\documentstyle[12pt]{article}
\newlength{\pubnumber} 

\catcode`\@=11
\@addtoreset{equation}{section}

\def\section{\@startsection{section}{1}{\z@}{3.5ex plus 1ex minus .2ex}
 {2.3ex plus .2ex}{\large\bf}}
\def\subsection{\@startsection{subsection}{2}{\z@}{2.3ex plus .2ex}
 {2.3ex plus .2ex}{\bf}}

\begin{document}

\begin{titlepage}
\samepage{
\setcounter{page}{1}
\rightline{\tt hep-ph/0104274}
\rightline{April 2001}
\vfill
\begin{center}
   {\Large \bf Solving the Hierarchy Problem without Supersymmetry or 
               Extra Dimensions: \\ An Alternative Approach\\ }
\vfill
   {\large
      Keith R. Dienes\footnote{
     E-mail address:  dienes@physics.arizona.edu}
    \\}
\vspace{.12in}
 {\it  Department of Physics, University of Arizona, Tucson, AZ  85721  USA\\}
\end{center}
\vfill
\begin{abstract}
  {\rm In this paper, we propose a possible new approach towards solving
   the gauge hierarchy problem without supersymmetry and without extra
   spacetime dimensions.  This approach relies on the finiteness of string
   theory and the conjectured stability of certain non-supersymmetric string
   vacua.  One crucial ingredient in this approach is the idea of 
   ``misaligned supersymmetry'', which explains how string theories
   may be finite even without exhibiting spacetime supersymmetry.
   This approach towards solving the gauge hierarchy problem is therefore
   complementary to recent proposals involving both large and small extra
   spacetime dimensions.
   This approach may also give a new perspective towards 
   simultaneously solving the cosmological constant problem.
   }
\end{abstract}
\vfill
\smallskip}
\end{titlepage}

\setcounter{footnote}{0}

\def\beq{\begin{equation}}
\def\eeq{\end{equation}}
\def\beqn{\begin{eqnarray}}
\def\eeqn{\end{eqnarray}}
\def\sosixteen{{$SO(16)\times SO(16)$}}
\def\V#1{{\bf V_{#1}}}
\def\half{{\textstyle{1\over 2}}}
\def\ttwo{{\vartheta_2}}
\def\tthree{{\vartheta_3}}
\def\tfour{{\vartheta_4}}
\def\ttwob{{\overline{\vartheta}_2}}
\def\tthreeb{{\overline{\vartheta}_3}}
\def\tfourb{{\overline{\vartheta}_4}}
\def\thetai{{\vartheta_i}}
\def\thetaj{{\vartheta_j}}
\def\thetak{{\vartheta_k}}
\def\thetaibar{\overline{\vartheta_i}}
\def\thetajbar{\overline{\vartheta_j}}
\def\thetakbar{\overline{\vartheta_k}}

\def\etainv{{\overline{\eta}}}
\def\Str{{{\rm Str}\,}}
\def\bone{{\bf 1}}
\def\chibar{{\overline{\chi}}}
\def\Jbar{{\overline{J}}}
\def\qbar{{\overline{q}}}
\def\calO{{\cal O}}
\def\calE{{\cal E}}
\def\calT{{\cal T}}
\def\calM{{\cal M}}
\def\calF{{\cal F}}
\def\calY{{\cal Y}}
\def\rep#1{{\bf {#1}}}
\def\ie{{\it i.e.}\/}
\def\eg{{\it e.g.}\/}

\newcommand{\newc}{\newcommand}
\newc{\gsim}{\lower.7ex\hbox{$\;\stackrel{\textstyle>}{\sim}\;$}}
\newc{\lsim}{\lower.7ex\hbox{$\;\stackrel{\textstyle<}{\sim}\;$}}

\hyphenation{su-per-sym-met-ric non-su-per-sym-met-ric}
\hyphenation{space-time-super-sym-met-ric}
\hyphenation{mod-u-lar mod-u-lar--in-var-i-ant}


\def\inbar{\,\vrule height1.5ex width.4pt depth0pt}

\def\IC{\relax\hbox{$\inbar\kern-.3em{\rm C}$}}
\def\IQ{\relax\hbox{$\inbar\kern-.3em{\rm Q}$}}
\def\IR{\relax{\rm I\kern-.18em R}}
 \font\cmss=cmss10 \font\cmsss=cmss10 at 7pt
\def\IZ{\relax\ifmmode\mathchoice
 {\hbox{\cmss Z\kern-.4em Z}}{\hbox{\cmss Z\kern-.4em Z}}
 {\lower.9pt\hbox{\cmsss Z\kern-.4em Z}}
 {\lower1.2pt\hbox{\cmsss Z\kern-.4em Z}}\else{\cmss Z\kern-.4em Z}\fi}

\def\NPB#1#2#3{{\it Nucl.\ Phys.}\/ {\bf B#1} (#2) #3}
\def\PLB#1#2#3{{\it Phys.\ Lett.}\/ {\bf B#1} (#2) #3}
\def\PRD#1#2#3{{\it Phys.\ Rev.}\/ {\bf D#1} (#2) #3}
\def\PRL#1#2#3{{\it Phys.\ Rev.\ Lett.}\/ {\bf #1} (#2) #3}
\def\PRT#1#2#3{{\it Phys.\ Rep.}\/ {\bf#1} (#2) #3}
\def\CMP#1#2#3{{\it Commun.\ Math.\ Phys.}\/ {\bf#1} (#2) #3}
\def\MODA#1#2#3{{\it Mod.\ Phys.\ Lett.}\/ {\bf A#1} (#2) #3}
\def\IJMP#1#2#3{{\it Int.\ J.\ Mod.\ Phys.}\/ {\bf A#1} (#2) #3}
\def\NUVC#1#2#3{{\it Nuovo Cimento}\/ {\bf #1A} (#2) #3}
\def\etal{{\it et al.\/}}

\long\def\@caption#1[#2]#3{\par\addcontentsline{\csname
  ext@#1\endcsname}{#1}{\protect\numberline{\csname
  the#1\endcsname}{\ignorespaces #2}}\begingroup
    \small
    \@parboxrestore
    \@makecaption{\csname fnum@#1\endcsname}{\ignorespaces #3}\par
  \endgroup}
\catcode`@=12

\input epsf

\section{Introduction}
\setcounter{footnote}{0}

In recent years, there has been much discussion of new methods of solving
the gauge hierarchy problem. 
In general, these approaches fall into several categories:
\begin{itemize}
\item  {\it Weak-scale supersymmetry}\/~\cite{susyreviews}:  
       This approach has the benefit of stabilizing
     the desert between  the electroweak symmetry-breaking scale
     and the fundamental high-energy scales such as the GUT scale,
     the Planck scale, and the string scale.
     Weak-scale supersymmetry also has many other virtues, among them
     the unification of gauge couplings, a possible triggering of electroweak
     symmetry breaking, and the prediction of various
     possible dark-matter candidates.
     Unfortunately, the paradigm of weak-scale supersymmetry is also
     fraught with a number of difficulties.
     For example, since supersymmetry must be broken, we face the  
     difficult question of how this breaking occurs and whether a hidden sector
     (and consequently a messenger sector) of some sort must be postulated 
     in order to accomplish this.
     A more serious difficulty, however, is that weak-scale supersymmetry
     must be broken in the TeV range.  Since this sets the scale for
     supersymmetry breaking, we cannot rely on weak-scale supersymmetry
     for a simultaneous solution to the cosmological constant problem.
\item  {\it Large Extra Spacetime Dimensions}\/~\cite{Antoniadis,extradims}:
     Over the past three years, it has become understood that large
     extra spacetime dimensions have the potential to lower the fundamental
     high-energy scales of physics, such as the 
     string scale, the Planck scale,
     and the GUT scale.
     As such, this approach towards solving the gauge hierarchy problem
     eliminates (rather than stabilizes) the desert, but in certain
     cases this occurs at the cost of reintroducing a desert in the
     required compactification radii.  Moreover, although weak-scale supersymmetry is
     no longer required for the purposes of stabilizing the hierarchy,
     supersymmetry of some sort ({\it e.g.}\/, in the bulk, or on distant
      branes) may still be required for the stability
     of the theory and of the brane configurations it requires.
     Most difficult, however, is the cosmological constant
     problem.  Once again, this problem is not automatically solved 
     simply by lowering the GUT, Planck, or string scales into the TeV range, 
      and  an additional mechanism still seems to be required.
\item  {\it ``Warped'' Spacetimes}\/~\cite{RS}:
     More recently, it has been shown that even {\it small}\/ extra spacetime
     dimensions can generate a large hierarchy between the Planck scale and
     the electroweak scale as a result of the warping of spacetime that occurs
     near a brane-like configuration of stress energy.
     Unlike the scenario involving large extra spacetime dimensions,
     this idea  has the virtue of generating the hierarchy in a natural way.
     However, this approach also faces a number of outstanding challenges.
     First, it is not clear whether or how the required preconditions for
     this scenario can be generated from string theory.  Second, it does
     not address (in and of itself) the cosmological constant problem;
     additional mechanisms of various sorts still seem to be required~\cite{recent}.
     Finally, supersymmetry may again be necessary in the full higher-dimensional
     theory in order to guarantee the fine-tunings of brane tensions that are
     required in this approach.
\item  {\it Conformality}\/~\cite{FV}:
     Finally, there exists another approach based on the 
      concept of ``conformality''.
     The basic idea is that new (as yet undiscovered) states may 
      exist in the TeV range such that,
     with these states included, the full theory becomes conformal 
     (scale-invariant) and has vanishing beta-functions, even while remaining
     non-supersymmetric.  Given this conformality, 
     the theory essentially experiences no ``running'' beyond the TeV scale;
     all scale-sensitivity is lost, 
     and the technical hierarchy problem is therefore solved.
     This approach is similar in spirit to that based on weak-scale supersymmetry:
     one simply replaces one symmetry (supersymmetry) with another (conformal symmetry)
     above the TeV scale, and both approaches (and indeed all of the above approaches)
     predict distinctly new physics near the TeV range. 
     However, once again, this approach faces a number of difficulties.
     First, one must have a way of breaking the conformal symmetry in the TeV range;
     this seems to be a particularly challenging issue given the presumed scale
     invariance of the theory.  Second, this approach does not shed any light
     on the cosmological constant problem;  just as with weak-scale supersymmetry, 
     one obtains a cosmological constant in the TeV range if the underlying
     symmetry is broken in the TeV range.  Finally, it has been pointed out~\cite{Csaki} 
     that within this scenario, even the hierarchy problem may not be completely solved.
\end{itemize}

The purpose of this paper is to propose an alternative approach towards
addressing the hierarchy and cosmological constant problems.
Unlike previous approaches, we will not employ either supersymmetry
or extra spacetime dimensions;  instead, our primary motivation
comes directly from studies of non-supersymmetric string theories 
and their physical spectra.
We also hasten to point out that we will not be completely 
solving the hierarchy problem (nor for that matter the 
cosmological constant problem), nor will we be presenting
a specific model that accomplishes all of these
tasks.  
Rather, our purpose is simply to present an alternative
philosophy towards thinking about these problems.

We also point out that for our purposes, the term ``gauge hierarchy
problem'' will be taken to refer to the technical question of the {\it stability}\/
of the electroweak scale rather than its origin.
Likewise, we will define the cosmological constant problem as the
problem of explaining why this constant vanishes (or at least is
insensitive to heavy scales)~\cite{Weinberg}.

\section{Developing an alternative approach:~  Four guiding\\ principles}
\setcounter{footnote}{0}

Our approach ultimately rests on four guiding principles which form the core
of the proposal.

\begin{itemize}
\item  Our first guiding principle is that {\it the hierarchy problem and
     cosmological constant problem are essentially the same problem}.
     Indeed, both are problems of stabilizing a light scale
     (either the electroweak symmetry-breaking scale or the scale
     of the cosmological constant itself) against the effects of a heavy
     fundamental scale ({\it e.g.}\/, $M_{\rm GUT}$, $M_{\rm Planck}$,
     or $M_{\rm string}$) with which it communicates quantum-mechanically.
     Given the commonality of these two problems, it is natural to expect
     that they should have a common, simultaneous solution.  Unfortunately, none of the
     standard approaches has this feature.  Although supersymmetry
     stabilizes the gauge hierarchy and also predicts a vanishing
     cosmological constant, the need to break supersymmetry in the TeV range
     leads to a cosmological constant which is also in the TeV range.
     The approach based on conformal symmetry also has a similar problem.
\end{itemize}

     Indeed, on general grounds, if a symmetry (such as supersymmetry) is to
     protect the hierarchy but must be broken in the TeV range, then any
     protection it provides for the cosmological constant is also lost
     in the TeV range.  This then leads to our second guiding principle:

\begin{itemize}
\item   We seek a symmetry {\it which stabilizes the gauge hierarchy,
    but which does not need to be broken}\/.  This implies that our symmetry
    should already be consistent with the Standard Model at low energies,
    which in turn implies that when unbroken, our symmetry must not intrinsically
    predict degenerate superpartners, conformal partners, or other 
    light states that are not seen.  
\end{itemize}

Our third guiding principle concerns the way in which we should think about energy
scales.  In field theory, we are encouraged to think of a linear ordering
of energy scales from lowest to highest, as if arranged along a line.  
Moving from high energy scales to low energy scales, 
the heavy states ``decouple'';  we can then integrate them out to obtain
effective theories at lower energies.  

String theory, by contrast,
teaches us something different:  near at the fundamental string scale,
we cannot necessarily distinguish heavy from light, small from large.
This has often been attributed to the ``quantum geometry'' of string theory.
The simplest example of this is the phenomenon of T-duality for closed
strings:  a closed string propagating in an extra spacetime dimension
compactified on a circle of radius $R$ has the same physics ({\it i.e.}\/, the same
physical spectrum,
the same scattering amplitudes, {\it etc.}\/) as it would have if
the spacetime were compactified
instead on a circle of radius $1/R$ (in units of $M_{\rm string}$).
This suggests that near the string scale,
we should abandon our usual field-theoretic notions by which we order
our energy scales in a linear fashion.
This observation becomes particularly relevant for phenomenology given
the fact that the fundamental string scale may be anywhere between
the Planck scale and the TeV scale. 

Indeed, if we look within string theory to see how     
the miracle of T-duality arises, we find that it is achieved through
a conspiracy between physics at {\it all energy scales simultaneously}.
One cannot ``integrate out'' the heavy string states (\eg, winding-mode states) 
while retaining light string states (\eg, Kaluza-Klein states),
since T-duality is achieved only through internal symmetries that relate
these states to each other throughout the string spectrum.
In other words, T-duality is intrinsically string-theoretic, and
does not survive into an effective field theory which might be generated
by integrating out heavy states.
We are therefore led to ask:

\begin{itemize}
\item Can we solve the hierarchy and cosmological
      constant problems this way, through a conspiracy 
      involving physics at all scales simultaneously?
      In such an approach, states at all energy scales should play an equal role;
      no states are to be ``integrated out''.
\end{itemize}
What we seek, therefore, is not a ``no-scales'' solution (as in the conformal approach),
but rather an ``all-scales'' solution, one in which states at all scales play an
equal role simultaneously.

Finally, again taking a cue from string theory, one of the intrinsic features
of string theory is the presence of an {\it infinite number of states in the string
spectrum}\/. 
We are referring here not only to string Kaluza-Klein states, but also to string winding
states (in the case of closed strings) as well as string resonances (\ie, excitations
of the string itself).  Like the Kaluza-Klein states, the latter  states also populate 
all mass scales simultaneously, but their numbers grow {\it exponentially}\/ as functions
of mass.  Together, these states are what set string
theory apart from field theory, giving string theory its remarkable finiteness
properties.

\begin{itemize}
\item Can we likewise exploit an {\it infinite}\/, exponentially growing set of states 
       at all mass scales in order to approach the hierarchy problem?  
\end{itemize}

Thus, to summarize, we seek an approach involving a symmetry
which has the following properties:  it need not be broken at any
scale (and hence must have no conflict with the 
Standard Model itself);  all mass scales should play an equal role in
conspiring to maintain this symmetry;  it should involve an infinite 
number of states, as suggested by string theory;
and it should be capable of addressing (if not solving) 
the hierarchy and cosmological constant problems simultaneously.

\section{Building a toy model}
\setcounter{footnote}{0}

The requirements discussed above clearly amount to a tall order, 
and it is not readily apparent how to proceed.
In order to develop some intuition, let us therefore  
begin by thinking about ordinary unbroken supersymmetry.
How does unbroken supersymmetry manage to solve the hierarchy and
cosmological constant problems?

In general, in supersymmetric theories, the quantum-mechanical sensitivities of light 
energy scales (such as the Higgs mass $m_H$ and 
the cosmological constant $\Lambda$) to heavy mass scales (\eg, a cutoff $\lambda$)
are governed by supertraces:
\beqn
     \delta m_H^2 &\sim& (\Str \calM^0) \lambda^2 + (\Str \calM^2) \log \lambda + ... \nonumber\\
     \Lambda &\sim&  (\Str \calM^0) \lambda^4 + (\Str \calM^2) \lambda^2 + (\Str \calM^4) \log \lambda ~ + ...
\label{divergences}
\eeqn
In these expressions, the ellipses `...' denote terms which are independent of the 
cutoff $\lambda$, and 
the supertraces are defined as statistics-weighted sums over the spectrum of the theory:
\beq
              \Str \calM^{2\beta} ~\equiv~ \sum_{{\rm states}~i} (-1)^F (M_i)^{2\beta}~.
\label{supertracedefFT}
\eeq
Thus, $\Str \calM^0$ (which merely counts the difference between the
numbers of bosonic and fermion states in the theory) 
governs the quadratic divergence in the one-loop
Higgs mass and the quartic
divergence in the one-loop cosmological constant,
while $\Str \calM^2$ governs the logarithmic divergence in the Higgs
mass and the quadratic divergence in the cosmological constant, and
$\Str \calM^4$ governs the logarithmic divergence in the cosmological constant. 
Of course, the superstraces relevant for the Higgs mass shift are to
be evaluated over only those states which couple to the Higgs;  by contrast,
the supertraces relevant for the cosmological constant are to be evaluated
over the complete spectrum of states in the theory.  

Supersymmetry works by ensuring that each of these supertraces vanishes.
Indeed, unbroken supersymmetry implies
\beq
      \Str \calM^{2\beta}~=~0  ~~~~~{\rm for~all}~~ \beta\geq 0~.  
\label{vanishingsupertraces}
\eeq
As a result of this feature,
the Higgs mass and the cosmological
constant each lose their quantum-mechanical sensitivities to heavy cutoff 
scales $\lambda$, and can remain light even after quantum-mechanical effects 
are included.
It is in this way that unbroken
supersymmetry solves the hierarchy and cosmological constant problems.
Indeed, this feature persists beyond one-loop, order by order.
Moreover, the cosmological constant not only loses its sensitivity to
heavy scales;  it actually vanishes. 
Thus, we see that the cancellations inherent in supersymmetry are 
encoded as cancellations of the mass supertraces.

Of course, in the real world, supersymmetry is broken.
However, even when the supersymmetry is broken, the technical gauge hierarchy problem remains
solved so long as
\beq
      \Str \calM^0 ~=~ 0~,~~~~~~ \Str \calM^2 ~\lsim~ ({\rm TeV})^2~.
\label{tevsutrace}
\eeq
Thus the cancellations inherent in supersymmetry 
continue to function in suppressing the mass supertraces, even when the
supersymmetry is softly broken.

This picture works well if we are concerned only with the gauge hierarchy
problem.  However, this picture becomes uncomfortable if we attempt to approach
the cosmological constant problem at the same time, for  the supertraces given in
(\ref{tevsutrace}) imply a cosmological constant which is also in the TeV range.
This is too large by many orders of magnitude.  From this perspective, one might say that 
the problem with supersymmetry is that its supertrace cancellations
occur multiplet-by-multiplet, scale-by-scale.  Thus, if supersymmetry
is broken at one scale, \eg, the TeV scale, as indicated in (\ref{tevsutrace}),
then this sets the same scale for the cosmological constant problem, destroying
any possible solution to both problems simultaneously.
The fundamental problem here is that {\it the scales for the gauge hierarchy
problem and the cosmological constant problem are very different}\/,
yet the breaking of supersymmetry sets a single scale for 
both problems simultaneously.

As indicated earlier, 
this suggests that we should not attempt to break our symmetry at {\it any}\/ scale.
Indeed, the only reason supersymmetry must be broken at all is that superpartners
are not detected in collider experiments at energies $E\lsim {\cal O}({\rm TeV})$;
this is what selects the TeV scale in (\ref{tevsutrace}).
This then leads to the fundamental question:
Is there a way to preserve (\ref{vanishingsupertraces}) through some other
symmetry which need not predict superpartners, and hence which would not need to be
broken at any scale?
In other words, is it possible to ensure the conditions (\ref{vanishingsupertraces})
for vanishing supertraces {\it without}\/ supersymmetry?

The conditions (\ref{vanishingsupertraces}) are very restrictive (since they
must be solved for all $\beta$ simultaneously), and for a finite number of states
it is easy to see that the only solution is to arrange all bosons and fermions
in the theory into exactly degenerate pairs.  This essentially restores the
supersymmetric configuration of the spectrum.
However, the key point is that we wish to consider theories with {\it infinite}\/
numbers of states.  Remarkably, it turns out that for infinite numbers of states,
other non-supersymmetric
boson/fermion configurations leading to vanishing supertraces are possible.

Before we can proceed, however, we must first generalize the definition of 
supertraces given in (\ref{supertracedefFT}).  While the definition (\ref{supertracedefFT})
is sufficient for theories containing finite numbers of states, it clearly has
the potential to diverge in theories containing infinite numbers of states.
We shall therefore introduce a regulated version of the supertrace:
\beq
          \Str \calM^{2\beta} ~\equiv~ \lim_{y\to 0} 
         \,\sum_{{\rm states}~i} (-1)^F (M_i)^{2\beta}\, e^{-y M_i^2}~.
\label{supertracedef}
\eeq
Here the quantity $y$ acts as a regulator which damps out the contributions
from heavy states with $M_i\gg y^{-1/2}$;  in the limit $y\to 0$, the contributions
of all states are included in the sum.
For a spectrum containing a finite number of states, this definition reduces to (\ref{supertracedefFT}),
while for a spectrum containing an infinite number of states, this definition will 
yield convergent results for all cases of interest.
At this stage, the form of this regular may seem completely arbitrary.
However, we shall see in Sect.~5 that
this form for the regulator is motivated by string-theoretic considerations,
and respects the underlying symmetry that we will eventually 
be proposing as a replacement for supersymmetry or conformal invariance.

Given the generalized definition (\ref{supertracedef}), we now return to 
our original question:  Is it possible to satisfy the vanishing supertrace 
constraints (\ref{vanishingsupertraces}) without supersymmetry?
As a warm-up exercise, let us first restrict our attention to the 
technical gauge hierarchy problem and ask whether we can satisfy
the simpler constraints 
\beq
       \Str \calM^0 ~=~0~,~~~~~~~~ \Str \calM^2 ~=~0~
\label{firsttwosupertraces}
\eeq
without supersymmetry.
As we shall see later, this will turn out to be sufficient for solving both
the technical hierarchy problem {\it as well as}\/ the cosmological constant problem.   

To do this, we shall construct our toy model by starting with a set of bosonic
states, and then balancing these with additional fermionic states as needed. 
Let us begin therefore with a simple set of two massless bosonic states.
(We choose two bosons rather than a single boson for future convenience.)
For our purposes, since we are merely counting degrees of freedom, we will not 
distinguish the spins of our states beyond noting
whether they are bosons or fermions. 

A non-supersymmetric configuration of two massless bosons clearly leads 
to quantum-mechanical divergences that destabilize the gauge hierarchy.
The standard resolution would be to introduce two new compensating
fermionic states with a mass $\mu$.  
Here $\mu$ functions as an arbitrary mass splitting.
We then have
\beq
         \Str \calM^0 ~=~0~,~~~~~\Str \calM^2 ~=~ -2\,\mu^2~.
\eeq
The vanishing of $\Str \calM^0$ ensures that the
quartic divergences are now cancelled.
Likewise, the value of $\mu$ is flexible and might be chosen, if we wish,
so as to satisfy external phenomenological criteria.
Of course, the only way to satisfy (\ref{firsttwosupertraces}) in this context
is to take $\mu=0$, yielding an exact boson/fermion degeneracy.

Let us now construct our toy model by considering how such a situation might be
extended to one with infinite numbers of states.
To do this, let us imagine duplicating the above double bosonic/fermionic
system infinitely many times at ever-increasing mass scales.
Specifically, we shall imagine that we have two states
for each mass $M_n=\sqrt{n}\mu$ for $n=0,1,2,...$.
If the level $n$ is even (respectively odd), we will assume that our states are
bosonic (respectively fermionic).
This particular form for the masses $M_n$
is motived by string theory and chosen for later convenience.
Likewise, we shall not specify a value for the overall mass scale
$\mu$ because we wish to
illustrate a general mechanism;  hence we shall not ``integrate out''
any heavy states.
Using the regulated supertrace defined in (\ref{supertracedef}),
we then find
\beqn
            \Str \calM^0 &=&
         2\,\lim_{y\to 0}\, \left\lbrace
         \sum_n \,(-1)^n \, e^{-y n \mu^2} \right\rbrace \nonumber\\
         &=& 2\,\lim_{y\to 0}\, \left(
              {1\over 1+ e^{-y \mu}} \right) \nonumber\\
         &=& 1~.
\label{exresultone}
\eeqn
Note that this formal result $\Str \calM^0=1$ reflects the expected averaging
between the values $+2$ and $0$ taken by $\Str \calM^0$ as the summation is performed
over the spectrum.
Likewise, we can also calculate $\Str \calM^2$ for our example:
\beqn
            \Str \calM^2 &=&
         2\,\lim_{y\to 0}\, \left\lbrace
         \sum_n \,(-1)^n \,n \mu^2\, e^{-y n \mu^2}
                                       \right\rbrace \nonumber\\
         &=& 2\,\lim_{y\to 0}\,
          {d\over dy}\, \left\lbrace
         \sum_n \,(-1)^n e^{-y n \mu^2}\right\rbrace
                                       \nonumber\\
         &=& 2\,\lim_{y\to 0}\,
          {d\over dy}\, \left(
          {1\over 1+ e^{-y \mu^2}}\right) \nonumber\\
         &=& {1\over 2}\, \mu^2~.
\label{exresulttwo}
\eeqn
Once again, the result is {\it finite}\/ even though infinitely many
states with infinite masses are present.

Given these results, it is quite easy to see how we might ensure that both
of our supertraces vanish.
Note from (\ref{exresultone}) and (\ref{exresulttwo}) that 
this infinite system of states has exactly the same supertraces as we would obtain
from a single bosonic state of mass $\mu/\sqrt{2}$.
Thus, all we need to do is introduce a single extra {\it fermionic}\/ state with mass
$\mu/\sqrt{2}$ (which we would associate with a level $n=1/2$).  
This combined configuration --- two repeating sets of pairs of
bosons/fermions at integer values of $n$, plus a single extra fermionic state 
at $n=1/2$ ---
has vanishing values for both $\Str \calM^0$ and $\Str \calM^2$ simultaneously.

Of course, this is nothing but a simple toy model, and we do not expect
this sort of configuration to be ``natural'' or to arise in any
well-motivated physical theory.  But it does illustrate several key features:
\begin{itemize}
\item  First, we see that
    {\it cancellation of divergences does not require a strict pairing of bosonic
   and fermionic states if there are infinite numbers of each}\/.
   In the above example, the presence of the single extra fermionic state
   destroys our na\"\i ve attempt at a pairing, yet is absolutely necessary
   in order to produce vanishing supertraces.
\item  Second, this result holds for all values of $\mu$;  no mass scale is preferred or selected
     by this procedure.
\item  Third, if we imagine that $\mu$ is large 
        (\eg, if $\mu\approx M_{\rm Planck}\sim 10^{18}$~GeV),
       then the only states
       that might be experimentally detected are those with masses $M_i\ll \mu$.
       In our toy model, these would just be the two single massless bosonic states
       with which we started --- a very non-supersymmetric light spectrum.
       However, even though we would not detect any massless or light ``superpartners''
       for these massless bosons,
       their divergences will nevertheless be cancelled
       by an infinite ``cloud'' of very heavy, experimentally
       undetectable particles with masses $M_n=\sqrt{n} \mu$,
       plus a fermion with mass $\mu/\sqrt{2}$.
       Note that a cancellation of this form requires
       an {\it infinite}\/ number of states in this theory.
       The actual value of the scale $\mu$ is irrelevant.
\item  Finally, we see that for the purposes of such calculations,
       we cannot integrate out our ``heavy'' states, even as $\mu\to \infty$.
       Instead, regardless of the value of $\mu$,
       keeping track of {\it all}\/ states in the theory is necessary in order to 
         analyze the full divergence
       structure of the low-energy states. 
       In other words, although a {\it finite}\/ number of heavy states can be expected
       to decouple as their masses become large, 
       an {\it infinite}\/ number of states will not.
\end{itemize}

Despite these features, the above toy model is quite unnatural for a number of reasons.
Perhaps its most unnatural feature is the {\it ad hoc}\/ introduction of the extra
fermionic state to cancel the supertraces.  By its very nature, this fermionic
state is not naturally part of our unified tower of states.
Another unnatural feature is our restriction to having only
two states for each mass $M_n$.  Once again taking a cue from string theory,
we expect that as the mass increases, the number of states
should also increase, for there are increasing numbers of ways in which
one can excite the fundamental string in order to produce increasingly
massive states.

Finally, another unnatural feature of our toy model is the fact
that we ``almost'' have an exact pairing of states --- indeed, except for
the solitary
extra fermionic state at mass $\mu/\sqrt{2}$,
all of the other states can easily be paired and have splittings that
scale with $\mu$.
This situation is very reminiscent of configurations with broken supersymmetry.
However, we are seeking scenarios in which there is no remnant of
multiplet-by-multiplet supersymmetry, approximate or otherwise.

Fortunately, it is not difficult to construct boson/fermion configurations
which overcome all of these difficulties.
For example, let us again consider a distribution of states
with masses $M_n=\sqrt{n}\mu$, bosonic if $n$ is even and
fermionic if $n$ is odd.
Let us further assume that the number of such states of
each mass $M_n$ is given by some number $g_n$,
with positive (negative) values of $g_n$ signifying bosons (fermions). 
Then solutions such as
\beq
   g_n ~=~ \cases{
            (-1)^n \,n^{2k} & for any $k\geq 1,\, k\in\IZ$ \cr
            (-1)^n \,(n^5-n) &  ~~\cr
            (-1)^n \,(n^5 + 2n^3) & ~~\cr}
\eeq
all have the property that they satisfy (\ref{vanishingsupertraces}).
Indeed, not only do $\Str \calM^0$ and $\Str \calM^2$ cancel identically,
but all higher mass supertraces cancel as well ---
all without any boson/fermion degeneracies!
For example, the case with $g_n=(-1)^n (n^5+2n^3)$ corresponds to an
explicit configuration of low-lying bosonic and fermionic states
consisting of
three fermions with mass $\mu$;
          48   bosons with mass $\sqrt{2}\mu$;
          297   fermions with mass $\sqrt{3}\mu$;
          1152   bosons with mass $ 2 \mu$;
          3375   fermions with mass $\sqrt{5} \mu$;
and so forth.
There is no apparent way to form a state-by-state pairing between
bosons and fermions, yet this configuration of bosons and fermions
manages to have vanishing supertraces of arbitrarily high order.
In particular, there is no supermultiplet structure, and no extraneous
counterbalancing states are required.
As in all cases, however,
the crucial ingredient in the success of this scenario
is the existence of an infinite number of states.
Indeed, it is only because of the subtle interplay between states
 {\it at all mass scales}\/ that the supertraces are cancelled
and the dependence on the scale $\mu$ is eliminated.

However, even these solutions are not ideal.
One glaring problem, for example, is the fact that no
massless states arise in these solutions;  all of these solutions have
$g_0=0$.  In the cases where $\mu$ is
large, we are therefore left with no light states.
Of course, 
we always have the freedom to superimpose onto this configuration
any other (supersymmetric) configuration with equal numbers of
bosons and fermions at any mass.
This does not disturb the vanishing supertrace relations that we have already
managed to satisfy.
However, even though this would introduce massless states into
the picture, our massless world would still be required to be supersymmetric.

Another (related) problem
is the fact that the degeneracies of states in these solutions
grow only polynomially in $n$.
If we wish to take the suggestions of string theory seriously,
then we should seek solutions for $g_n$ which grow {\it exponentially}\/
according to the asymptotic formula
\beq
         |g_n| ~\sim~ A n^{-B} e^{C\sqrt{n}} ~~~~~{\rm as}~ n\to \infty~
\label{Hagedorn}
\eeq
where $A$, $B$, and $C$ are all positive constants.
This is the expected asymptotic behavior for the number of ways of producing
a state consisting of $n$ total excitations using only integer
excitation modes.

It turns out to be much more difficult to construct
non-trivial solutions that also satisfy these two additional criteria.
However, using techniques from string theory,
solutions can be found.
In this section, we shall simply present what we shall refer to as a ``magic'' solution.
In Sect.~5, we shall discuss in detail how such solutions may be generated
and interpreted.

Our ``magic'' solution consists of the following low-lying degeneracies $g_n$:
\begin{table}[h]
\centerline{
   \begin{tabular}{||c|c||c|c||}
   \hline
   \hline
    $n$  &  $g_n$  & $n$  &  $g_n$  \\
   \hline
   \hline
    $0$  &  $+36$    &   $6$  &   $ -29,010,432$ \\
    $1$  &  $+1,024$    &   $7$  &   $-29,774,848$ \\
    $2$  &  $-19,712$    &   $8$  &   $+529,050,944$ \\
    $3$  &  $-76,800$    &   $9$  &   $+410,305,536$ \\
    $4$  &  $+1,051,136$    &   $10$  &   $-7,301,403,648$ \\
    $5$  &  $+1,806,336$    &   $11$  &   $-4,414,798,848$ \\
   \hline
   \hline
   \end{tabular}
}
\end{table}

\noindent Thus, unlike the previous cases, we see that this solution has bosons
at levels $n\in \lbrace 0,1\rbrace $~(mod~4) 
and fermions at levels $n\in\lbrace 2,3\rbrace$~(mod~4).
Of course, in order to fully specify this solution, we need to quote
more than a finite set of numbers $g_n$;  we need to specify a procedure by which {\it all}\/
values of $g_n$ can be generated.  This procedure will be given
explicitly in Sect.~5.

This ``magic'' solution has a number of appealing properties.
First, it is explicitly non-supersymmetric;  there exists no manifest way of pairing
states at any mass level in order to construct supersymmetric multiplets.
Moreover, since $g_0\not=0$, we see that the lightest states are themselves 
non-supersymmetric;  thus even as $\mu\to \infty$ we retain a non-supersymmetric
spectrum of light states.
Furthermore, these degeneracies $g_n$ exhibit\footnote{
         More precisely, 
         each separate sequence of integers $|g_n|$ with a different value of $n$~(mod 4)
         obeys (\ref{Hagedorn}) individually.
         This is sufficient for our purposes, and 
         will be discussed further in Sect.~4.}
the desired exponential growth anticipated in (\ref{Hagedorn}).

But most importantly, our ``magic'' solution has two critical properties.
First, both of the supertraces relevant for the gauge hierarchy problem vanish
identically:
\beq
         \Str \calM^0 ~=~ \Str \calM^2 ~=~ 0~.
\eeq
These supertrace cancellations occur as a result of a non-trivial conspiracy
across the infinite particle spectrum in this theory.
Thus, because of these cancellations, the gauge hierarchy 
ceases to depend on the splitting scale $\mu$ or any other high fundamental 
scale. 

Second, it turns out that the one-loop cosmological constant corresponding
to this solution actually vanishes.
Recall that for an arbitrary set of degeneracies $g_n$, 
the one-loop cosmological constant (or equivalently 
the one-loop zero-point vacuum energy) is given by
\beq
    \Lambda ~=~ \half \sum_n \, g_n\,
                \int {d^4 p \over (2\pi)^4} \,\log(p^2+M_n^2)~.
\label{Lambdadef}
\eeq
The dependence on the degeneracies $g_n$ is shown explicitly in this equation.
However, as we shall see in Sect.~5, for this particular choice
of $g_n$ it is possible to extract a finite result for $\Lambda$ in 
a self-consistent way, and moreover, this finite result actually 
vanishes.  Thus, for the particular choice of $g_n$ in this ``magic''
solution, we have not only vanishing supertraces but also $\Lambda=0$.
Note that this has been achieved {\it without}\/ supersymmetry
and without any explicit boson/fermion pairings.
Moreover, this holds for all values of the splitting parameter $\mu$.

What are the phenomenological implications of such a scenario?
One possibility is to take $\mu\approx M_{\rm Planck}\sim 10^{18}$~GeV
and to identify the ``massless'' states as those of the Standard Model.
Of course, the Standard Model does not consist merely of 36 bosonic degrees of freedom.  
However, given one solution with the above properties, it is trivial to
generate others.  For example, all degeneracies $g_n$ may be subjected to an arbitrary
common rescaling, since this does not affect the vanishing of the
supertraces.  This implies, in particular, that we can
also reverse the signs of all of the degeneracies, thus exchanging bosons
and fermions without affecting the final result.  Moreover, at each mass level,
we are free to add an arbitrary {\it equal}\/ number of bosons and fermions,
since  $g_n$ represents only the {\it net}\/ number of bosons minus fermions at the $n^{\rm th}$
mass level.  Given these facts, it is always possible to accommodate the Standard Model
states within the lightest states in such a configuration. 

We therefore see that it is possible to think of the Standard Model states as being
only the lightest states in a theory containing an infinite number of states.
For sufficiently large mass splittings $\mu$, no other light states need appear,
yet by properly summing over the infinite spectrum of such a theory, one finds
that all supertraces cancel.   Our low-energy theory consequently loses 
its quantum-mechanical dependence on heavy mass scales, 
and the {\it technical}\/ hierarchy ``problem''    
is reduced to a fictitious problem that appears only because we 
have integrated out an {\it infinite}\/ tower of states.
As we have seen, the cosmological constant problem
may also be addressed simultaneously
in this approach.

Clearly, this scenario is radically different from the usual
scenario involving weak-scale supersymmetry.
Rather than embellish the Standard Model particles with weak-scale
superpartners, we are instead embellishing them with a ``cloud'' of infinitely many heavy
states which perform essentially the same function, namely that of
regulating and/or cancelling divergences.
Thus, we need not impose supersymmetry
and then worry about how to break it --- instead, we never have it at all.
Moreover, in this scenario, the new particles are not necessarily
at the weak scale;  the precise scale depends on the (arbitrary) value
of $\mu$.  However, because the configuration of these heavy
particles has been very carefully chosen, a conspiracy between physics
at all energy scales occurs,
and the quantum-mechanical dependence of the low-energy theory 
on heavy mass scales is eliminated.

\section{Connection to string theory}
\setcounter{footnote}{0}

In some sense, what we have so far is merely a mathematical 
result concerning a new way to cancel supertraces.
But the deeper question remains:  why should one believe that
this is relevant for the real world?

Fortunately, the best motivation for this approach is that
 {\it this is what string theory actually does}.
Recall that string theory is a finite theory:  the finiteness 
of string theory rests not on supersymmetry, but rather on the fact that
strings are extended objects.
It is this extended nature of the string which provides an automatic
``regulator'' for the ultraviolet (short-distance) divergences that normally plague
quantum field theories.  If the string model in question happens to exhibit
spacetime supersymmetry, then this finiteness manifests itself in the usual
manner, via  pairwise cancellations throughout the string spectrum.  
However, even if the string is non-supersymmetric, this finiteness must be
maintained.\footnote{We are here ignoring a number of subtle issues concerning
the stability of non-supersymmetric strings.  We will discuss these issues
more carefully below.} 
In such cases, {\it the mechanism we
have presented in the previous section is the method by which string theory
manages to adjust its spectrum to maintain finiteness, even without supersymmetry}.  
Rather than rely on 
the level-by-level cancellations inherent in supersymmetry,
the cancellations leading to finiteness 
in this scenario instead occur non-trivially across the entire 
string spectrum in the manner just described.
In fact, this sort of cancellation has been proven~\cite{missusy,supertraces} 
to be a general property of all closed, perturbative,
tachyon-free non-supersymmetric strings.

As an example, let us consider what is perhaps the most famous perturbative
non-supersymmetric string theory:  this is the ten-dimensional tachyon-free
heterotic string model with gauge group $SO(16)\times SO(16)$, originally constructed
in Ref.~\cite{so16model}.
The spectrum of this model has alternating, exponentially increasing
bosonic and fermionic surpluses:
\beq  (g_0;\,g_1;\,g_2;\,...) ~=~  (-2,112; ~ 147,456; ~ -4,713,984; ~...)~.
\eeq
Moreover, as we shall discuss in Sect.~5, one finds that the first {\it four}\/
mass supertraces all vanish in this model~\cite{supertraces}: 
\beq
        \Str \calM^0 ~=~  \Str \calM^2~=~\Str \calM^4~ =~ \Str \calM^6~=~0~.
\label{so16sutraces}
\eeq

As shown in Refs.~\cite{missusy,supertraces},
the underlying symmetry which assures that these cancellations take place is
nothing but modular invariance.  This is hardly surprising, since
modular invariance has its roots in the extended nature of the string,
and serves as a powerful constraint on the string spectrum. 
Thus,  modular invariance (and its multiloop generalizations) 
is the symmetry that we are proposing to replace
supersymmetry or conformal symmetry  as our solution to the hierarchy problem. 
Indeed, modular invariance is a general property 
of the perturbative moduli space of all closed string theories
(and the bulk sector of Type~I string theories).
However, unlike supersymmetry or conformal symmetry, {\it there is no need
to break modular invariance for phenomenological purposes}.
Thus, our symmetry and the cancellations it induces can remain intact
even at low energies.

This also answers another question.  {\it A priori}\/, the cancellation
outlined in the previous section might have seemed to be a highly fragile one, 
since  a small shift in the mass of a single state (\eg, due to a radiative
correction) would seem to destroy the
delicate supertrace cancellations we have managed to achieve.
However, these cancellations are ultimately robust because they are the manifestation
of deeper symmetry, in this case modular invariance.
Thus, if one state in the string spectrum is shifted due to some dynamical effect
calculated within the framework of the full string theory,
the rest of the string spectrum automatically compensates in a modular-invariant
way.  Thus, the supertrace cancellations are ultimately preserved and protected 
by modular invariance.

\begin{figure}
\vskip -1.0 truein
\centerline{
   \epsfxsize 6.0 truein \epsfbox {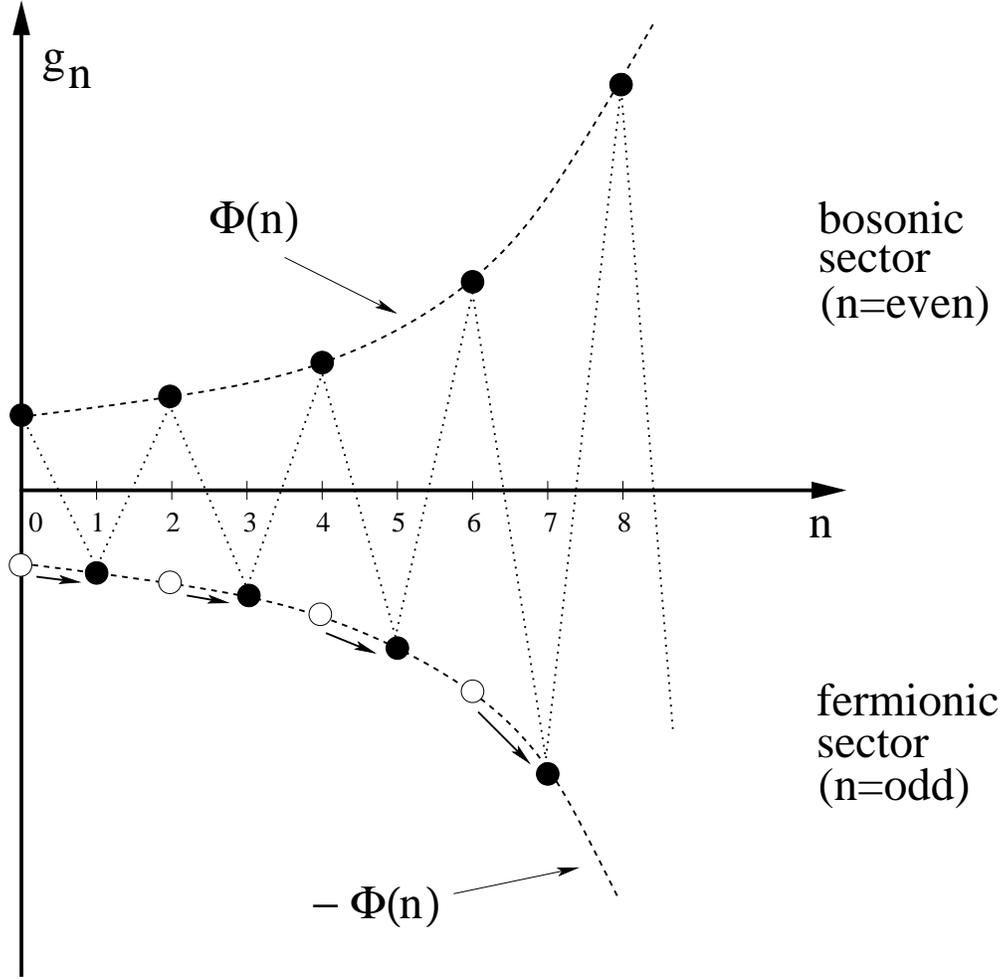}
    }
\vskip -1.0 truein
\caption{A sketch of a typical boson/fermion configuration.
In this sketch, we have assumed a bosonic (fermionic) sector at even (odd) mass levels $n$,
and plotted a typical configuration of degeneracies $g_n$ (black circles) versus $n$.
The dashed lines connect these points in order of increasing mass $M_n=\sqrt{n}\mu$, 
and illustrate the regular bosonic/fermionic oscillations inherent 
in ``misaligned supersymmetry''.
The cancellation of the supertraces arises as a result 
of the {\it cancellation of the functional forms}\/ $\Phi(n)$ that 
separately govern the behavior of $g_n$ as a function of $n$ in each sector. 
Ordinary supersymmetry emerges as a special case
when the bosonic and fermionic sectors have values of $n$ that coincide
(no misalignment).
We illustrate this for even values of $n$, with degeneracies (black circles) in the bosonic sector
cancelling pairwise against degeneracies (white circles) in the fermionic sector.
However, as the fermionic sector is shifted (``misaligned'') by $n\to n+\Delta n$ relative
to the bosonic sector, the white circles slide along the functional form $-\Phi(n)$
to their new locations $-\Phi(n+\Delta n)$ at which pairwise cancellations of states 
are no longer possible.  Although supersymmetry is broken, finiteness is nevertheless 
maintained due to the cancellation of the functional forms, 
and the mass supertraces continue to vanish.}  
\label{missusyfig}
\end{figure}

It is easy to understand explicitly how finiteness is achieved
in such alternating boson/fermion scenarios.
For this purpose, let us consider a typical boson/fermion configuration,
as sketched in Fig.~\ref{missusyfig}.
As we scan through higher and higher mass levels, the spectrum oscillates
between exponentially growing bosonic and fermionic surpluses.
In the sketch in Fig.~\ref{missusyfig}, for example, the sectors with
bosonic surpluses occur at mass levels where $n$ is even, while
the sectors with fermionic surpluses occur when $n$ is odd.
Clearly, in such a configuration, there is no pairwise cancellation
of states.  Nevertheless, in each sector separately, there exists a
unique way~\cite{HR,missusy} of determining a
smooth function $\Phi(n)$ which describes the exponentially growing 
degeneracies $g_n$ at the appropriate values of $n$.  For large $n$,
these functions typically take the form 
\beq
      \Phi(n) ~=~ \sum_{\ell=1}^\infty \sum_i  
                 \, f(n+n_i)\, \exp\left( {c_i\over \ell}\sqrt{ n+ n_i} \right)~ 
\label{Phiform}
\eeq
where $\lbrace c_1,c_2,c_3,...\rbrace$ are a set of positive real
numbers;  where $0\leq n_i< 1$ for all $i$;
and where $f(x)$ is a function which grows at most polynomially in $x$.
Thus, $\Phi(n)$ takes the form of an infinite series of
leading and subleading exponentials, with the 
leading term, as expected, taking the Hagedorn form (\ref{Hagedorn}) for
large $n$.
Given this, the cancellation that preserves
finiteness in such configurations is not a cancellation of the degeneracies
$g_n$, but rather a cancellation of the functional forms $\Phi(n)$ between
the different sectors of the theory~\cite{missusy}.  Specifically, if $\Phi_b(n)$
represents the functional form describing the degeneracies $g_n$ in the bosonic
sector and $\Phi_f(n)$ represents the analogous functional form for $|g_n|$ in the 
fermionic sector, then
\beq
                { \Phi_b(n) - \Phi_f(n) \over
                  \Phi_b(n) + \Phi_f(n) } ~\to~0~~~~~~~~ {\rm as}~~ n\to\infty~.
\label{partialone}
\eeq
In other words, the leading (and often also the highest subleading) exponential 
terms in (\ref{Phiform}) necessarily
cancel between bosonic and fermionic sectors.
In fact, for {\it any}\/ non-supersymmetric tachyon-free closed string model with
multiple bosonic and fermionic sectors, it has been shown~\cite{missusy} that
\beq
        { \sum_i    \,(-1)^{F_i}  \, \Phi_i(n) \over \sum_i \,\Phi_i(n)}~\to~0~~~~~~~~ 
               {\rm as}~~ n\to\infty~ 
\label{partialtwo}
\eeq
where $F_i$ is the fermion number for the $i\/^{\rm th}$ sector.
This is thus a completely general property.
Moreover, it has been conjectured~\cite{missusy} that above cancellations
even take the stronger form
\beq
         \sum_i    \,(-1)^{F_i}  \,\Phi_i(n) ~=~0~.
\label{conjecture}
\eeq
This would correspond to a complete cancellation of {\it all}\/ terms within
(\ref{Phiform}).
We will discuss these results in more detail below.

In some sense, this mechanism is a generalization of the level-by-level pairing
of states that occurs in ordinary unbroken supersymmetry.
To see this, let us return to Fig.~\ref{missusyfig}
and imagine that the bosonic and fermionic sectors were exactly aligned,
so that both had mass levels only for even values of $n$.
In this case, the degeneracies $g_n$ for the bosonic sector would be
exactly aligned with those for the fermionic sector, and the cancellation of
functional forms in (\ref{conjecture})
would imply an exact pairwise cancellation of states.  Indeed, this is what happens
for a supersymmetric string model.  However, as we break supersymmetry and
shift the mass levels of the fermionic sector relative to those in the bosonic sector
by an amount $\Delta n$,
the degeneracies $g_n$ in the fermionic sector shift according to\footnote{
         Strictly speaking, as supersymmetry is broken, 
         the relevant values of $n$ shift but the
         functional forms $\Phi(n)$ also change in such a way as to ensure that
         the corresponding degeneracies $g_n$ are always integers.} 
\beq       g_n= -\Phi_f(n)~~~\longrightarrow~~~  g_n' = -\Phi_f(n+\Delta n)~.
\eeq
This shift is illustrated in Fig.~\ref{missusyfig}.
In other words, the fermionic states 
redistribute themselves across the entire infinite string spectrum 
in such a way that although the level-by-level pairing of states is destroyed,
the cancellation of the functional forms is still preserved.
This can therefore be called a ``misaligned supersymmetry''~\cite{missusy}.
Thus, ``misaligned supersymmetry'' and its associated cancellations 
is the underlying mechanism by which 
modular invariance is maintained in the string spectrum, even 
when supersymmetry is no longer present.
In fact, as shown in Ref.~\cite{missusy},
this the most general way by which supersymmetry can be broken in string theory while
simultaneously maintaining modular invariance (a critical symmetry for the perturbative
consistency of closed strings) and avoiding the appearance of tachyonic string states.

Note that this mechanism goes beyond a mere uniform shifting of the masses of
certain states.  While some states become heavier because of the misalignment,
other states (typically winding-mode states) rearrange themselves and become lighter 
such that the net values 
of $g_n$ continue to fall along
the same functional form $\Phi_f(n)$.  Such a rearrangement is possible only in a theory
with an {\it infinite}\/ number of states.  

\begin{figure}    
\centerline{
   \epsfxsize 3.3 truein \epsfbox {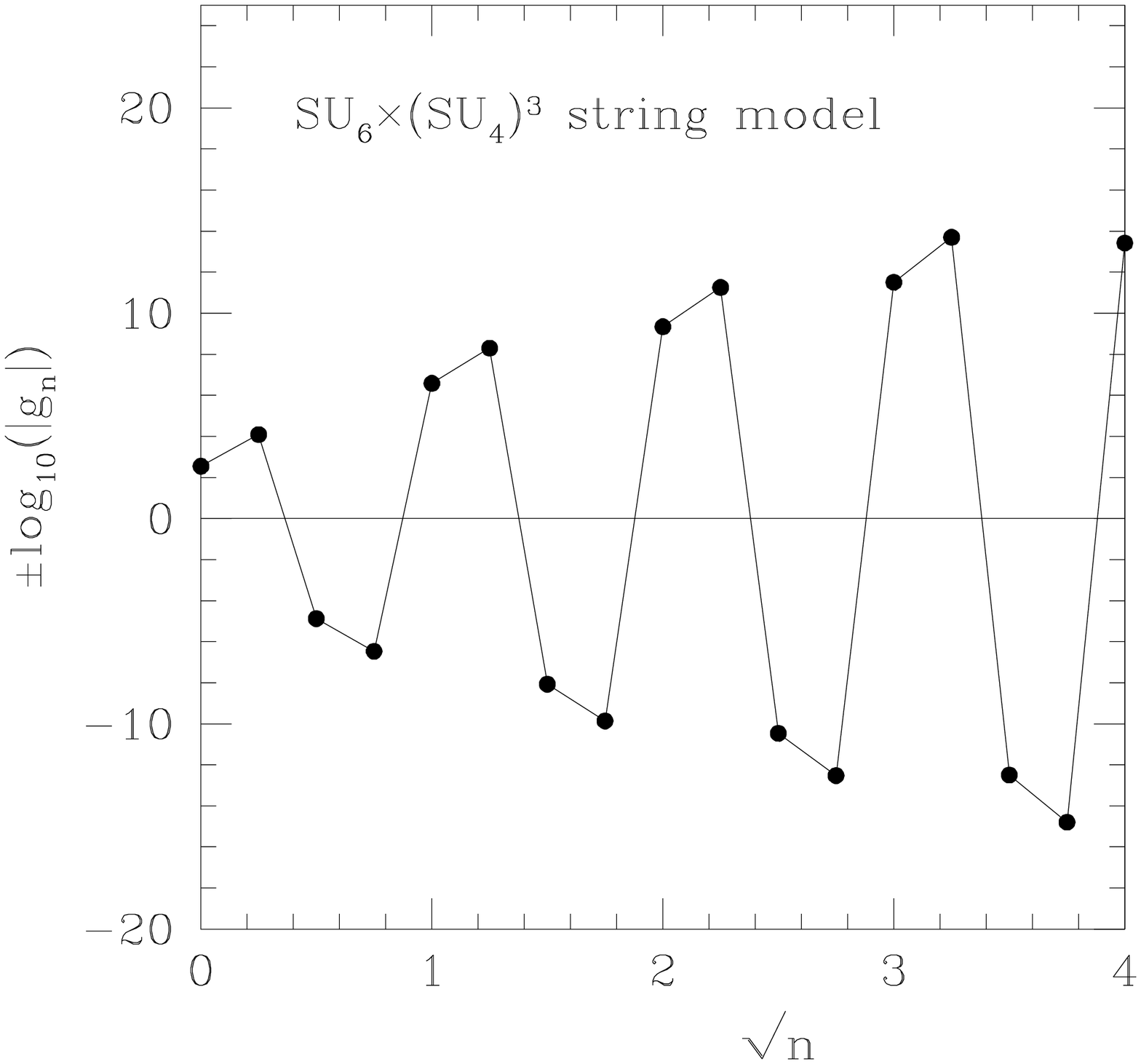}
   \epsfxsize 3.3 truein \epsfbox {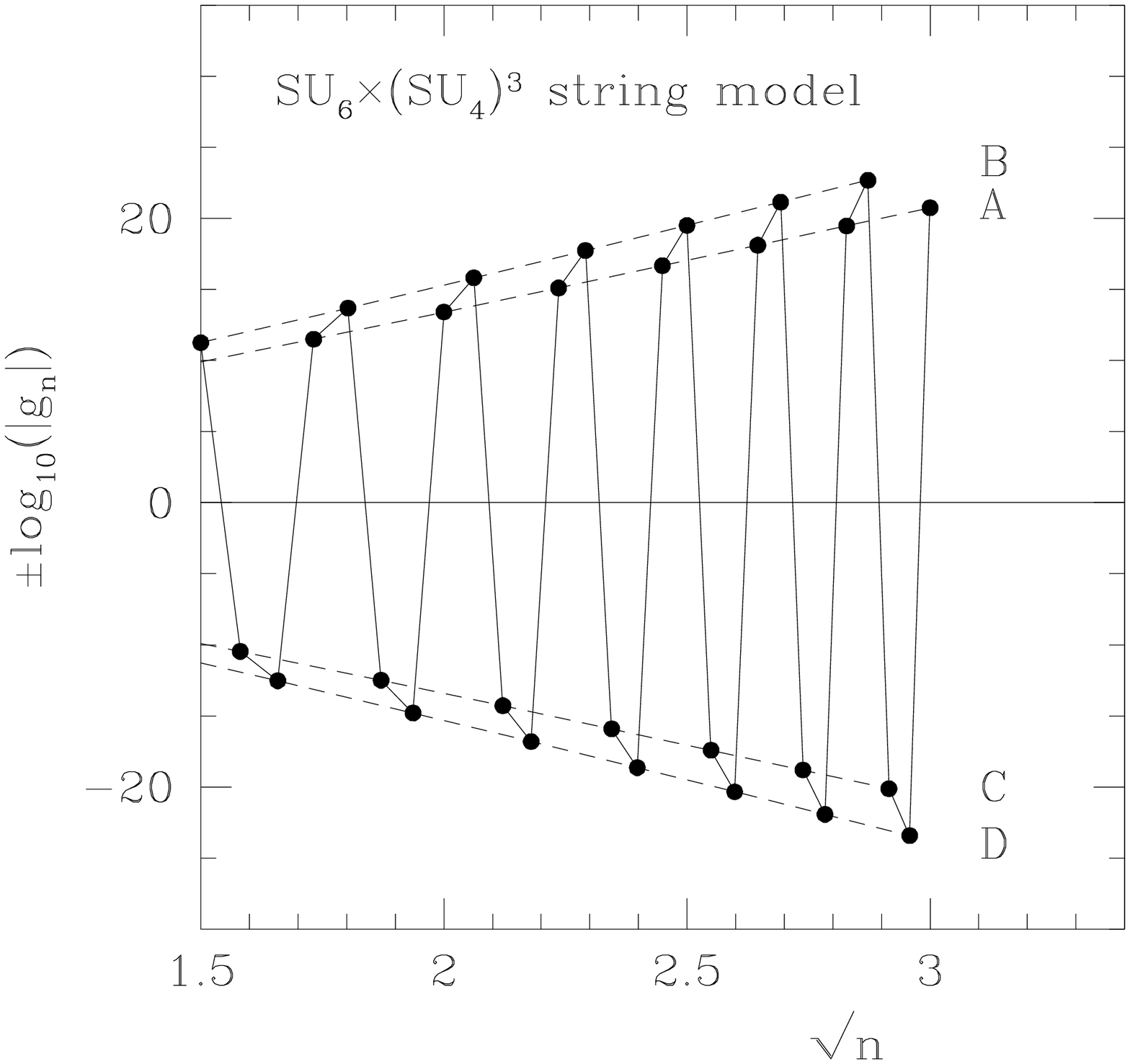}
    }
\centerline{
   \epsfxsize 3.3  truein \epsfbox {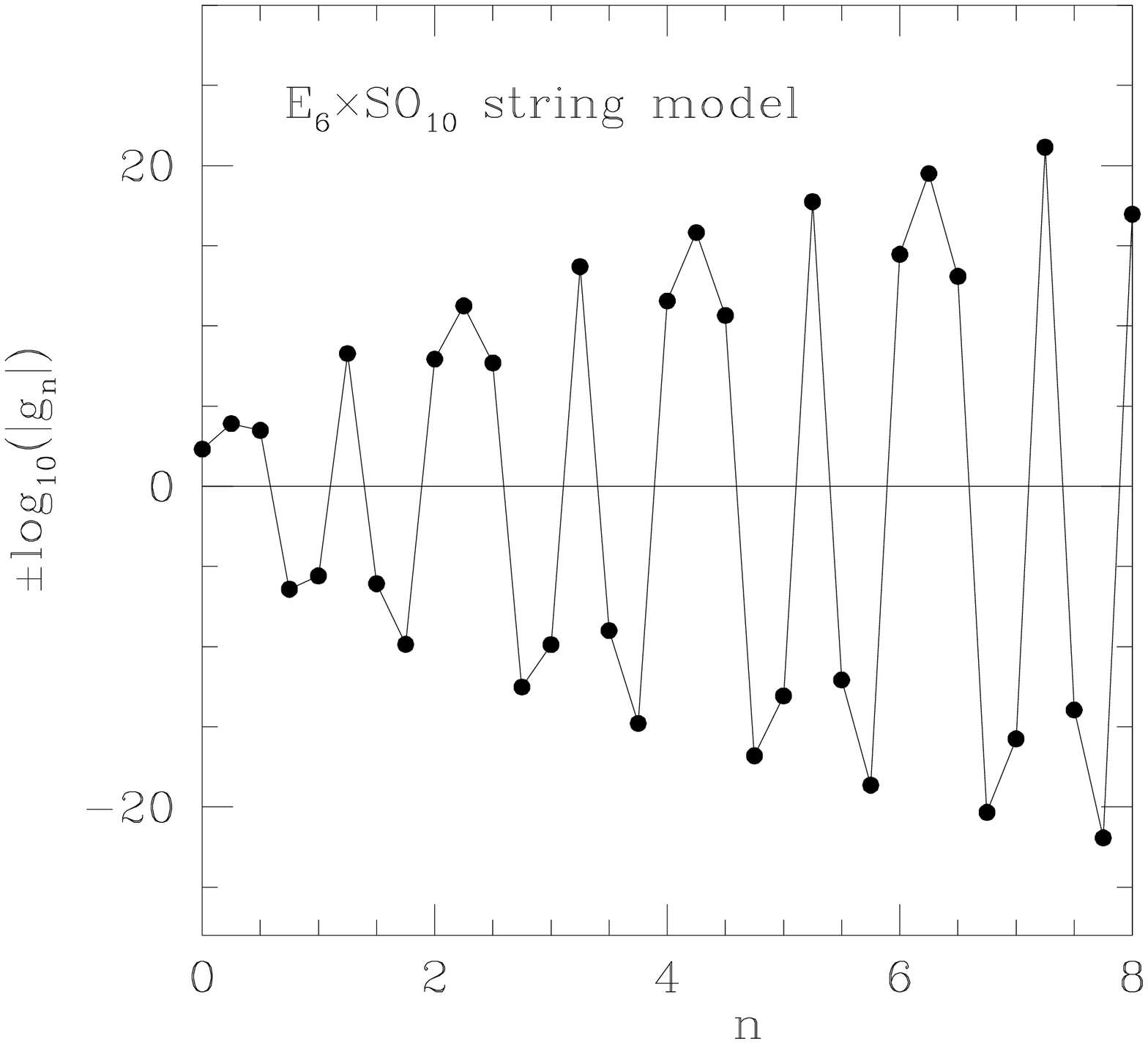}
   \epsfxsize 3.3  truein \epsfbox {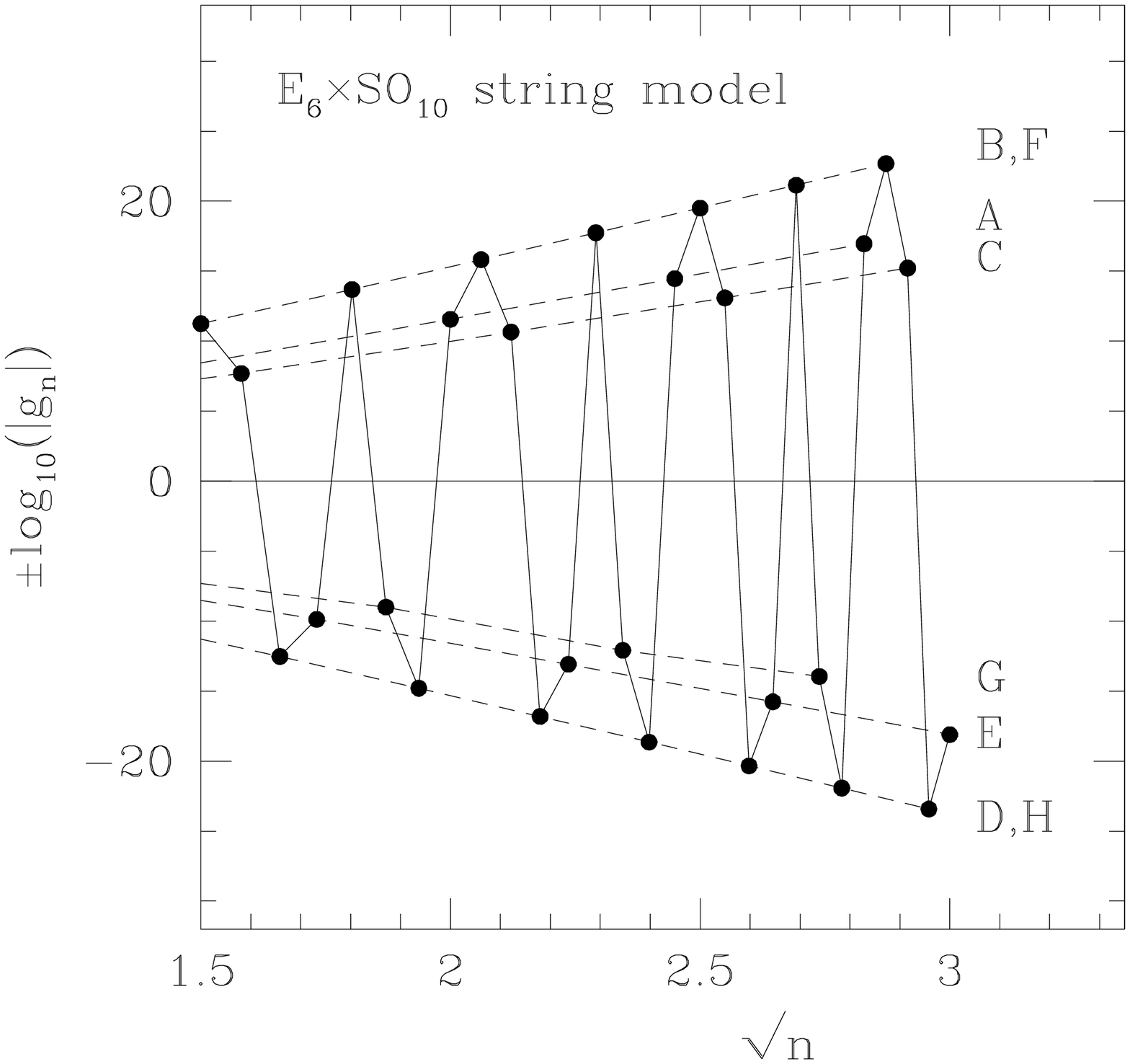}
    }
\caption{ Degeneracies $g_n$ as functions of $n$ in four-dimensional
      non-supersymmetric tachyon-free heterotic string models with gauge groups 
         $SU_6 \times (SU_4)^3$ (top row) and $E_6\times SO_{10}$ (bottom row).  
        In these figures we plot $\pm \log_{10}(|g_n|)$ where the sign chosen is the
        sign of $g_n$.  In all cases, modular invariance causes cancellations 
       to occur, leading to a misaligned supersymmetry and preserving finiteness.}     
\label{exampleplots}
\end{figure}

Of course, Fig.~\ref{missusyfig} is only a sketch of an idealized situation.
In actual string models, these cancellations can be far more complicated.
Let us consider two examples.
Our first example is a four-dimensional, non-supersymmetric, tachyon-free 
perturbative heterotic string model with gauge group $SU_6\times (SU_4)^3$.
(The full gauge group of this model is $SU_6 \times (SU_4)^3 \times (SU_2)^9 \times U_1$.)
This model is constructed via the heterotic free-fermionic construction~\cite{KLT}.
For this string model, we can plot the degeneracies $g_n$ as a function of $n$.
We then obtain the result shown in the top row of Fig.~\ref{exampleplots}.
Note that this model contains four sectors distributed
along quarter-integer values of $n$;  
those with $n\in \lbrace 0,1/4\rbrace$ (mod~1) have bosonic surpluses, 
while those with $n\in \lbrace 1/2,3/4\rbrace$ (mod~1) have fermionic
surpluses.  Labelling these sectors as $\lbrace A,B,C,D\rbrace$ respectively,
we see that there are four distinct functions $\Phi_{A,B,C,D}(n)$ which
describe the degeneracies in these sectors respectively.  
As indicated in Fig.~\ref{exampleplots} and as expected from (\ref{Hagedorn}), 
these functions are essentially linear on a log plot of $g_n$ versus $\sqrt{n}$.
Indeed, as evident from Fig.~\ref{exampleplots},
$\Phi_{B,D}$ have a stronger leading exponential behavior
than $\Phi_{A,C}$.
Nevertheless, it follows that
\beq
           {  \Phi_A(n) + \Phi_B(n) - \Phi_C(n) -\Phi_D(n) \over
    \Phi_A(n) + \Phi_B(n) + \Phi_C(n) +\Phi_D(n) } ~\to ~ 0 ~~~~~~~~~{\rm as}~~n\to \infty~, 
\eeq
with the conjectured stronger cancellation
\beq
             \Phi_A(n) + \Phi_B(n) - \Phi_C(n) -\Phi_D(n) ~=~ 0~.
\eeq
This implies that the leading exponential behavior between
$\Phi_B$ and $\Phi_D$ cancels exactly;  that the remaining
subleading exponential behavior from $\Phi_{B,D}$ cancels 
against the leading exponential behavior from $\Phi_{A,C}$;
and so forth.

As a more complicated example, let us consider a different string 
model with gauge symmetry $E_6\times SO_{10}\times (SU_4)^3\times (U_1)^2$.  
The degeneracies for
this model are plotted in the lower row of Fig.~\ref{exampleplots}. 
Unlike the previous model, this model has {\it eight}\/ distinct sectors
with values of $n\in \lbrace 0,1/4,1/2,3/4,1,5/4,3/2,7/4\rbrace$ (mod~2).
It is immediately apparent that the oscillation pattern in this model is
much more complex than it is in the previous model.  Nevertheless, labelling
these sectors $\lbrace A,...,H\rbrace$ respectively, we still find the 
net cancellation
\beq
        {\Phi_A +\Phi_B +\Phi_C - \Phi_D -\Phi_E +\Phi_F -\Phi_G -\Phi_H \over
        \Phi_A +\Phi_B +\Phi_C + \Phi_D +\Phi_E +\Phi_F +\Phi_G +\Phi_H }~\to~ 0~
           ~~~~~~~~~ {\rm as}~~n \to\infty~.
\eeq
Indeed, regardless of the particular model or its method of construction, 
modular invariance causes cancellations to occur, 
leading to a misaligned supersymmetry and preserving finiteness.     
Moreover, in each case, it is the presence of an infinite number of states
which permits such cancellations to occur.

\section{Technical details}
\setcounter{footnote}{0}

We shall now provide some technical background behind the
results which have been presented thus far.
This section is self-contained, and the reader uninterested in these
details can proceed directly to the Conclusions.
The purpose of this section is to provide a summary of those results
in the literature which provide further details behind the claims above. 

Our goal is to understand the emergence of modular invariance and 
its relation to misaligned supersymmetry, the mass supertraces, and the 
cosmological constant.
Let us therefore begin by considering the field-theoretic one-loop cosmological 
constant.  In $D$ dimensions, this is given by
\beq
     \Lambda ~=~ \half \sum_n \,(-1)^{F_n}\, g_n\,
                \int {d^D p \over (2\pi)^D} \,\log(p^2+M_n^2)~.
\label{LambdadefD}
\eeq
For $D=4$, this reduces to the expression given in (\ref{Lambdadef}).  
Clearly, this field-theoretic
expression has a divergence as $p^2\to \infty$.  
In order to understand how the presence of an infinite number
of string states enables us to eliminate this divergence,
let us follow the standard prescription by rewriting this expression in terms of a Schwinger
proper time $t$ by using the identity 
\beq
       \log\,x
     ~=~ \int_1^x \,{dy \over y}
     ~=~ \int_1^x dy\,\int_0^\infty dt ~ e^{-yt}
  ~=~ -\int_0^\infty {dt\over t}~e^{-xt}~+~...
\eeq
where we have dropped an $x$-independent term.
Since we are interested in the divergence behavior as
$x\to \infty$, it is legitimate to drop this term.
We thus obtain
\beq
   \Lambda~=~ -\half \sum_n \,(-1)^{F_n}\,g_n\,
  \int {d^D p \over (2\pi)^D}
      \,\int_0^\infty {dt\over t} e^{-(p^2+M_n^2)t}~,
\label{Lambdaschwinger}
\eeq
and the ultraviolet divergence as $p^2\to\infty$ now appears as
a divergence as $t\to 0$.
Performing the momentum integrations then yields
\beq
   \Lambda~=~ -{1\over 2} \,{1\over (4\pi)^{D/2}}\, 
      \sum_n \,(-1)^{F_n}\, g_n\,
  \int_0^\infty {dt\over t^{1+D/2}} e^{-M_n^2 t}~.
\label{nomom}
\eeq

In order to show how modular invariance can arise,
we now shift our notation slightly by making the following
substitutions.  First, we define
the {\it dimensionless}\/ real parameter $\tau_2$ as
\beq
       \tau_2 ~=~ {1\over 4 \pi}\, \mu^2 \,t
\eeq
where $\mu$ is an (as yet) unspecified mass scale.
Second, we introduce an additional
dimensionless real variable $\tau_1$ by inserting
\beq
        1 ~=~ \int_{-1/2}^{1/2} d\tau_1~
\eeq
into our expressions.
We then combine our two new parameters to form the complex variable
\beq
       \tau ~\equiv~ \tau_1 + i\tau_2 ~,
\eeq
thereby enabling us to rewrite our expression for $\Lambda$ in the form
\beq
     \Lambda~=~  -{1\over 2}\, \left({\mu\over 4\pi}\right)^D\,
     \int_{\cal S} {d^2\tau\over {\tau_2}^2}\, Z(\tau)
\label{Lambdafinal}
\eeq
where ${\cal S}$ denotes the semi-infinite strip in the complex $\tau$-plane
\beq
    {\cal S} ~\equiv~ \left\lbrace \tau~:~ |{\rm Re}\,\tau| \leq \half,\,
          {\rm Im}\,\tau \geq 0~\right\rbrace
\label{stripdef}
\eeq
and where the integrand $Z(\tau)$ is
\beqn
          Z(\tau) &\equiv& {\tau_2}^{1-D/2}\, \sum_n \,(-1)^{F_n} \,g_n\,
          \exp(-4\pi \tau_2 M_n^2/\mu^2)~ \nonumber\\
          &=&  {\tau_2}^{1-D/2}\,\sum_n \,(-1)^{F_n} \,g_n\,
               (\overline{q}q)^{M_n^2/\mu^2}~
\label{Zphys}
\eeqn
with $q\equiv e^{2\pi i\tau}$.

As our final step, let us now demand that $M_n= \sqrt{n} \mu$.
This assumption is valid
for all of the boson/fermion configurations we have given previously,
and also holds in string theory (thereby
permitting us to identify $\mu\equiv 2 M_{\rm string}$ using the standard
normalizations~\cite{GSW} for closed strings).
Given this assumption, we 
can then add any term of the form
\beq
               {\tau_2}^{1-D/2}\,
          \sum_{ \scriptstyle m\not =n\atop \scriptstyle m=n \,({\rm mod}\, 1)}
            \,g_{mn}~
         \overline{q}^{M_m^2/\mu^2} \, q^{M_n^2 /\mu^2}
\label{Zunphys}
\eeq
to $Z(\tau)$ without changing the value of the cosmological constant.
This is because any extra terms of the form (\ref{Zunphys})
integrate to zero across the strip (\ref{stripdef}),
regardless of the values of $g_{mn}$.
We can therefore combine (\ref{Zphys}) and (\ref{Zunphys}) in order to
write our total integrand as
\beq
              Z(\tau) ~=~  {\tau_2}^{1-D/2}\,\sum_{m,n}\, g_{mn}
         \,\overline{q}^{M_m^2/\mu^2} \, q^{M_n^2 /\mu^2}
\label{Ztot}
\eeq
where we have defined $g_{nn}= (-1)^{F_n} g_n$.
Note that in a supersymmetric field theory, all values of $g_{nn}$ vanish,
and hence $Z(\tau)$ and $\Lambda$ themselves also vanish.
Also note that with or without supersymmetry, $Z(\tau)$ has the
property that $Z(\tau) = Z(\tau+1)$.  In other words,
$Z(\tau)$ is invariant under
\beq
             \tau ~\to~\tau~+~1~.
\label{Tdef}
\eeq

As we know, the field-theoretic cosmological constant is usually
ultraviolet-divergent in the absence of supersymmetry.
We have already remarked that in the Schwinger-time formalism,
this ultraviolet divergence at $p^2\to\infty$ has been mapped to
a divergence as $t\to 0$
(or equivalently as $\tau$ approaches the real axis in the 
complex $\tau$-plane).
The fact that our strip integration region (\ref{stripdef}) includes
a portion of this axis is the reflection of this field-theoretic divergence.
The conventional method for removing this divergence
in field theory without supersymmetry is to introduce an {\it ad hoc}\/
ultraviolet cutoff $\lambda$\/ --- {\it i.e.}\/,
to restrict our momentum integrations
to the regions $p^2  \leq  \lambda^2$.
This cutoff procedure thereby introduces a new scale $\lambda$ into our theory
(independent of $\mu$), and
$\lambda$ then serves as the high scale that sets
the size of the resulting cosmological
constant and likewise destabilizes the gauge hierarchy.
In the language of the strip, one is essentially truncating the
integration region so that ${\rm Im}\,\tau\equiv \tau_2 \geq \lambda$
for some $\lambda >0$.  In this way, the real axis is excluded.

Is there another way of truncating the integration region which
essentially {\it avoids}\/ introducing a new cutoff scale?
Indeed, such a prescription is well known in string theory.
In string theory, the coefficients $g_{nn}$ correspond to the degeneracies
of physical string states, while  the off-diagonal coefficients $g_{mn}$ 
correspond to the degeneracies of so-called
``off-shell'' or unphysical string states.
Like ghosts in field theory, these states propagate
only in loops and do not exist as {\it bona-fide}\/ in-states or out-states.
However, in string theory
the physical and unphysical string states together conspire
to produce values of $g_{mn}$ in (\ref{Ztot}) such that
$Z(\tau)$ acquires the additional property 
that $Z(\tau) =  Z( -1/\tau)$.
In other words, $Z(\tau)$ is invariant not only under (\ref{Tdef}), 
but also under the additional complex
transformation
\beq
               \tau~\to~ -1/\tau~.
\label{Sdef}
\eeq
Moreover, since the measure of integration
$d^2\tau / {\tau_2}^2$
in (\ref{Lambdafinal}) is also invariant under the
transformations (\ref{Tdef}) and (\ref{Sdef}),
we see that these two transformations are a symmetry
of the entire amplitude, analogous to a gauge invariance.

Of course, the symmetry generated by the transformations
(\ref{Tdef}) and (\ref{Sdef}) is nothing but modular invariance,
and $Z(\tau)$ is the modular-invariant partition function of the theory.
However, just as with a gauge symmetry, 
when calculating a scattering amplitude
we must avoid overcounting by dividing out by the infinite symmetry volume
factor;  in other words, we must tally only those contributions which are 
inequivalent with respect to the symmetry.
We must therefore truncate our
strip region of integration so that the new (smaller) region of integration includes 
only one representative
value of $\tau$ up to the combined modular transformations (\ref{Tdef}) and (\ref{Sdef}).  
Such a region is given by
\beq
    {\cal F} ~\equiv~ \left\lbrace \tau~:~ |{\rm Re}\,\tau| \leq \half,\,
          |\tau| \geq 1 \right\rbrace~,
\label{Fdef}
\eeq
and is commonly called the {\it fundamental domain of the modular group}.
By excluding the real axis completely, modular invariance thus succeeds in cancelling
the ultraviolet divergence in the cosmological constant without introducing a new
fundamental cutoff scale beyond $\mu$;  essentially the modular symmmetry renders 
the divergence spurious  by enabling it to be reinterpreted as the infinite volume 
associated with a symmetry group.  Dividing out by this volume,   
one thus obtains a new, manifestly finite expression for the
cosmological constant:
\beq
     \Lambda~=~  -{1\over 2}\, \left({\mu\over 4\pi}\right)^D\,
      \int_{\cal F} {d^2\tau\over {\tau_2}^2}\, Z(\tau)~.
\label{Lambdafinalstring}
\eeq
Indeed, our discussion thus far has merely described the standard 
``recipe'' by which one calculates the 
one-loop cosmological constant in string theory~\cite{Polchinski}.

We have thus established that any distribution of states
which causes $Z(\tau)$ to be modular-invariant will
completely cancel the ultraviolet divergence of
the cosmological constant.
The crux of the matter, then, is to construct such modular-invariant
distributions.
It is clear that the case of unbroken supersymmetry manages to do
this trivially by forcing all $g_{nn}=0$;  indeed, string theories with
spacetime supersymmetry actually have $g_{mn}=0$ for {\it all}\/ $(m,n)$, so that
$Z(\tau)=0$ for all values of $\tau$.
This is a level-by-level, scale-by-scale cancellation.
However, 
our goal has been to find alternative solutions which
do this in a far less trivial manner.
Fortunately, we see that modular invariance does not require that
$Z(\tau)=0$.  Thus, as long as supersymmetry is broken {\it in such a way that
modular invariance is preserved}\/,
the divergence in the cosmological constant can be eliminated as outlined above.

What, then, is the most general way of breaking supersymmetry while preserving
modular invariance?  As discussed in Sect.~4, the supersymmetry can be at
most {\it misaligned}\/.
Indeed, as shown in Ref.~\cite{missusy},
the oscillating properties of misaligned supersymmetry discussed 
in Sect.~4 are nothing more than the residual constraints on the degeneracies
$g_{nn}$ that follow from demanding a 
modular-invariant partition function $Z(\tau)$.
These reflect themselves as a cancellation of the function forms $\Phi(n)$,
as discussed previously.

Given these results, it might seem that one requires knowledge of both the
physical ($m=n$) and unphysical ($m\not= n$) states 
in order to calculate the cosmological constant.  In particular,
the truncation of the region of integration from the strip ${\cal S}$
to the fundamental domain ${\cal F}$ in going from 
(\ref{Lambdafinal}) to (\ref{Lambdafinalstring})
implies that the unphysical
states now make a non-zero contribution to $\Lambda$. 
However, the physical and unphysical states are related to each
other via modular invariance, and it turns out that modular invariance
enables the total contribution to the cosmological constant 
from the unphysical states to be determined directly 
from just the physical states.
This implies that it should be possible to express our final 
expression for $\Lambda$ in (\ref{Lambdafinalstring}) directly in terms
of only the diagonal elements $g_{nn}$,
and it has been shown~\cite{KS} that such an expression 
is given by
\beqn
           \Lambda &=&  -{\pi\over 6}\, \left({\mu\over 4\pi}\right)^D\, 
                     \lim_{\tau_2 \to 0} \, \int_{-1/2}^{1/2} 
              d\tau_1 \, Z(\tau)\nonumber\\
            &=&  -{\pi \over 6}\, \left({\mu\over 4\pi}\right)^D\, 
                 \lim_{\tau_2 \to 0} \, {\tau_2}^{1-D/2}\,
                    \sum_n  \,g_{nn}\, \exp\left( -4\pi \tau_2 M_n^2 /\mu^2 \right)~.
\label{KSresult}
\eeqn
This expression applies for a large class of tachyon-free string theories
including all unitary non-critical strings, critical Type~II strings,
and heterotic strings in $D>2$.  Moreover, no spacetime supersymmetry is required.
Since $\Lambda$ has already been rendered finite by modular invariance,  
this implies that the physical-state degeneracies $g_{nn}$ in such modular-invariant 
string theories must have the property that as $\tau_2\to 0$, 
\beq
          \sum_n \, g_{nn} \, \exp\left(- 4\pi \tau_2 M_n^2/\mu^2 \right) 
            ~\sim~ -{6\over \pi}\left( {4\pi\over \mu}\right)^D \Lambda\, 
                 {\tau_2}^{D/2 -1}~+~...~, 
\label{limitingbehavior}
\eeq
where `...' refers to terms that vanish more rapidly than $\tau_2^{D/2-1}$ in the
$\tau_2\to 0$ limit. 

Given this result,
we can now calculate the 
mass supertraces over the physical string states
in modular-invariant theories in even 
numbers of spacetime dimensions~\cite{supertraces}. 
In terms of the net degeneracies $g_{nn}$, we have
\beqn
          \Str \calM^{2\beta} 
        &=& \lim_{y\to 0} \,\sum_n \, g_{nn} \, (M_n)^{2\beta}\, e^{-y M_n^2}~\nonumber\\
        &=& \lim_{y\to 0} \left\lbrace \left( -{d\over dy}\right)^\beta\,
                       \sum_n \, g_{nn} \,  e^{-y M_n^2}\right\rbrace~.
\eeqn
However, in the $y\to 0$ limit, we can use the result (\ref{limitingbehavior})
to evaluate the summation over string states:
\beq
          \Str \calM^{2\beta} 
        ~=~ \lim_{y\to 0} \, \left\lbrace \left( -{d\over dy}\right)^\beta\,
             \left\lbrack -{96\pi} \,{\Lambda\over \mu^2}\,  \left( 4\pi y\right)^{D/2-1} 
               + ...\right\rbrack\right\rbrace~.
\eeq
Consequently, we see that $\Str \calM^{2\beta}$ must necessarily {\it vanish}\/ for 
all $0\leq \beta < D/2-1$:
\beq
         \Str \calM^0 ~=~ \Str \calM^2 ~=~ ...~=~\Str \calM^{D-4}~=~0~.
\label{conc1}
\eeq
Moreover, for even spacetime dimensions, we see that the first non-zero supertrace is given
by
\beq
          \Str \calM^{D-2}~=~  24 \, (-4\pi)^{D/2} \, (D/2-1)! ~ {\Lambda\over \mu^2}~.
\label{conc2}
\eeq
Thus, the value of the first non-zero supertrace is 
set by the one-loop cosmological constant!

For $D=10$, these results imply that $\Str \calM^0$, $\Str \calM^2$, $\Str \calM^4$, and $\Str \calM^6$ all
vanish, even if supersymmetry is not present.
This is the origin of the result (\ref{so16sutraces}) claimed earlier.
On the other hand, for $D=4$, these relations reduce to
\beq
            \Str \calM^0 ~=~ 0~,~~~~~~~~~
          \Str \calM^{2}~=~  24 \, (4\pi)^{2} ~{\Lambda\over \mu^2}~.
\label{sutracerels}
\eeq
In each case, these relations are {\it guaranteed}\/ for any modular-invariant theory
regardless of its method of construction, its compactification manifold, or its
low-energy phenomenology.
No supersymmetry is required, softly broken or otherwise. 

As discussed in Ref.~\cite{supertraces},
these results may seem extremely counter-intuitive
from the point of view of ordinary four-dimensional field theory.
After all, these results imply that we can relate the actual value of
the one-loop amplitude $\Lambda$ to a single (mass)$^2$ supertrace. 
By contrast, in ordinary quantum field theory,
a supertrace of this order should describe
the {\it quadratic divergence}\/ of such an amplitude, not its constant term.
Indeed, in ordinary quantum field theory, we know that
$\Str \calM^0$, $\Str \calM^2$, and $\Str \calM^4$ respectively govern
quartic, quadratic, and logarithmic divergences in $\Lambda$.
The point here, however,
is that modular invariance is so powerful a symmetry that
it effectively softens the divergences of amplitudes by four
powers of mass.
Thus, in a theory with modular invariance,
the quartic and quadratic divergences are automatically cancelled ---
even without supersymmetry.
Moreover, the highest remaining divergence for such an amplitude is
a logarithmic one, now governed by  $\Str \calM^0$.
The vanishing $\Str \calM^0$ in these scenarios thus guarantees
the finiteness of $\Lambda$, and $\Str \calM^2$ then describes the
constant term --- \ie, the value of $\Lambda$ itself.
Moreover, as a byproduct, these results also verify that  
the regulator chosen for our supertraces
in (\ref{supertracedef}) respects modular invariance, as originally claimed.

For our purposes, however, the most important observation is
that {\it the gauge hierarchy problem is now related to the cosmological constant
problem at a fundamental level}\/.  
Indeed, the only way to ensure that both $\Str \calM^0$ and $\Str \calM^2$ vanish 
is to ensure that $\Lambda=0$.
Thus, at least at the one-loop level, {\it modular invariance and misaligned supersymmetry
tie these two fundamental
problems together}\/ in precisely the way we envisioned in Sect.~2.

Moreover, in string theory, a non-zero one-loop cosmological constant signifies an unstable
string vacuum.  This occurs because a non-zero one-loop zero-point function (cosmological constant) 
necessarily implies a non-zero one-loop dilaton tadpole.  
Thus, in string theory, the cosmological constant and gauge hierarchy
problems each become
tantamount to the third problem of vacuum stability!
In other words, they are related to the moduli problem.
Of course, the relation of the cosmological constant problem to the string moduli problem
is not new.  Our point here, however, is 
that in string theory, the problems of vacuum stability, cosmological constant, and gauge hierarchy
are now seen to be merely {\it one}\/ problem, not three separate problems.
Thus, any stable non-supersymmetric string will {\it automatically}\/
incorporate solutions to the remaining problems as well.  

We have demonstrated these claims merely to one-loop order
for strings in a flat background spacetime.  
However, it is likely that multi-loop generalizations of modular symmetry 
will guarantee the continuation of these connections to all orders, and perhaps 
even non-perturbatively.  Indeed, just like misaligned supersymmetry itself,
it is likely that such the connection between stability, the cosmological constant,
and the gauge hierarchy supertraces
is part of the consistency
of non-supersymmetric strings.

Needless to say, all of these results hold in a trivial manner even 
when an exact supersymmetry 
is present.  In the presence of unbroken spacetime supersymmetry, 
the string vacuum is automatically
stable (living along a flat potential in all directions), and the technical 
gauge hierarchy and cosmological constant problems are automatically solved as a 
result of this flatness.
Our claim, then, is that a similar situation persists even when the supersymmetry is broken:
as long as a stable non-supersymmetric string vacuum exists, that stability 
will manifest itself through the appearance of a misaligned supersymmetry,
and this misaligned supersymmetry
will be sufficient to guarantee a solution to the technical gauge hierarchy 
and cosmological constant problems.

Given these observations, the next step is clearly to find a stable non-supersymmetric 
string, or equivalently to find a string theory with $\Lambda=0$.  
Moreover, for our purposes, we wish to demand that such a string not even exhibit
boson/fermion degeneracies (so that $g_{nn}\not= 0$).
While the existence of such a string theory is not known at present,
we wish to offer the following observation.
In Sect.~4, we introduced two four-dimensional non-supersymmetric
tachyon-free heterotic string models, one with 
gauge group $SU_6\times (SU_4)^3 \times (SU_2)^9 \times U_1$
and the other with gauge group
$E_6 \times SO_{10} \times (SU_4)^3\times (U_1)^2$.
Even though these string models have unequal gauge groups, unequal particle contents,
and unequal bosonic and fermionic degeneracies at all mass levels,
it turns out that they manage to have exactly the same one-loop cosmological constant!
In other words, their one-loop cosmological constants are exactly equal, 
\beq
    \Lambda_{SU_6\times (SU_4)^3} ~=~ \Lambda_{E_6\times SO_{10}}~,
\eeq
even though their partition functions are unequal, 
\beq
    Z_{SU_6\times (SU_4)^3} ~\not=~ Z_{E_6\times SO_{10}}~.
\eeq
This sort of non-trivial degeneracy between non-supersymmetric string 
models was originally noticed in Ref.~\cite{firstpaper},
and occurs quite frequently in the moduli space of non-supersymmetric string theories.
Given this, it is possible to consider the {\it difference}\/ of partition functions 
\beq
       \Delta Z ~\equiv~  {1\over 4}\, 
         \left[ Z_{SU_6\times (SU_4)^3} ~-~ Z_{E_6\times SO_{10}}\right]
\eeq
in order to generate solutions $\lbrace g_{nn}\rbrace$ which have the property that $\Lambda=0$.
Specifically, given the partition functions of these string models, we find
\beqn
      \Delta Z &=&  {1\over 128 \,\tau_2}\, {1\over \overline{\eta}^{12}\, \eta^{24}} ~\times\nonumber\\
      && ~~~~\times \, 
       \sum_{\scriptstyle i,j,k=2  \atop \scriptstyle i\not= j\not= k}^4
      \,  |\thetai|^4 \,  \Biggl\lbrace  \,
 \thetai^4 \thetaj^4 \thetak^4
      \,\biggl\lbrack \,2\, |\thetaj \thetak|^8  -
     \thetaj^8 \thetakbar^{8} - \thetajbar^{8} \thetak^8  \biggr\rbrack
                    \nonumber\\
        &&~~~~~~~~~~~~~~ +~ \thetai^{12} \,\biggl[
      \, 4 \,\thetai^8  \thetajbar^{4} \thetakbar^{4} +
               (-1)^i~13 \,|\thetaj  \thetak|^8 \biggr] \, \Biggr\rbrace~ 
\label{DeltaZ}
\eeqn
where $\eta$ and $\vartheta_i$ are the Dedekind eta-function and Jacobi theta-functions:
\beqn
    \eta(q)  ~\equiv&  q^{1/24}~ \displaystyle\prod_{n=1}^\infty ~(1-q^n)&=~
                \sum_{n=-\infty}^\infty ~(-1)^n\, q^{3(n-1/6)^2/2}\nonumber\\
    \vartheta_2(q)~\equiv&  2 q^{1/8} \displaystyle\prod_{n=1}^\infty (1+q^n)^2 (1-q^n)&=~
                 2\sum_{n=0}^\infty q^{(n+1/2)^2/2} \nonumber\\
    \vartheta_3(q)~\equiv&  \displaystyle\prod_{n=1}^\infty (1+q^{n-1/2})^2 (1-q^n) &=~
                1+ 2\sum_{n=1}^\infty q^{n^2/2} \nonumber\\
    \vartheta_4(q) ~\equiv& \displaystyle\prod_{n=1}^\infty (1-q^{n-1/2})^2 (1-q^n) &=~
                1+ 2\sum_{n=1}^\infty (-1)^n q^{n^2/2} ~.
\eeqn
We can then do a power expansion of $\Delta Z$, 
\beq
       \Delta Z(q,\overline{q}) ~=~ {1\over \tau_2}\, \sum_{m,n} \,g_{mn} \, \overline{q}^m q^n~,
\eeq
and thereby generate a set of physical-state degeneracies $\lbrace g_{nn}\rbrace$
for which the corresponding cosmological constant vanishes identically:
\beq
          \Lambda ~=~ \int_{\cal F} \, {d^2\tau\over {\tau_2}^2}\, \Delta Z(\tau,\overline{\tau})~=~ 0~.
\label{ALzero}
\eeq
Moreover, upon rescaling $n\to 4n$, we find that these 
degeneracies $\lbrace g_{nn}\rbrace$ are precisely 
those of the ``magic'' solution presented in Sect.~3,
for which both $\Str \calM^0$ and $\Str \calM^2$ vanish identically!

Of course, the difference $\Delta Z$ of two string partition functions is not necessarily another
string partition function.  In other words, the set of integers $\lbrace g_{nn}\rbrace $ need 
not necessarily (and in this case does not) emerge as the state degeneracies of a stable 
non-supersymmetric string.
Despite this fact, this solution for $\lbrace g_{nn}\rbrace $ continues to have 
vanishing supertraces and
corresponds to a vanishing cosmological constant in the manner described above.
This solution is therefore perfectly valid from a purely field-theoretic point of view.

For this reason,
it is instructive to ascertain how these $\lbrace g_{nn} \rbrace$ manage to 
achieve a vanishing cosmological constant.  It turns out that the underlying mechanism 
rests on
the fact that the function $\Delta Z$ in (\ref{DeltaZ}) exhibits a 
so-called Atkin-Lehner symmetry~\cite{Moore}.
Specifically, for such a function $\Delta Z$, it is possible to rewrite 
the integral in (\ref{ALzero}) in such a way that the new integrand is odd under
a discrete {\it non}\/-modular transformation such as $\tau \to -1/2\tau$ while 
the new integration measure and domain are even under this transformation.  
The resulting integral thus vanishes as a result of a discrete selection rule.
Indeed, such Atkin-Lehner symmetries
are possible only in theories with infinite numbers of states, and once again
such cancellations emerge as the result of delicate conspiracies 
between physics at all mass scales.
Unfortunately, several attempts~\cite{attempts} and ultimately a no-go theorem~\cite{BT} 
have shown that no self-consistent string model
can have a partition function exhibiting such an Atkin-Lehner symmetry.
While a proposal has been made~\cite{GAL} for generalizing the idea of Atkin-Lehner symmetry 
to evade this no-go theorem,
no stable self-consistent non-supersymmetric string model 
has yet been constructed along these lines. 
 
More recently, it has been shown~\cite{KachSilv} that there exist
non-supersymmetric compactifications of Type~II strings for which
$\Lambda=0$ to one- and two-loop orders.
Moreover, it has been conjectured~\cite{KachSilv,KachSilvothers} that this cancellation
persists to higher loops as well, and perhaps even non-perturbatively.
As such, this is an exciting development.
However, even though these models are non-supersymmetric, they
nevertheless exhibit an exact level-by-level boson/fermion degeneracy.
They therefore have $g_{nn}=0$ for all $n$.  Thus,  
even though they are non-supersymmetric, they are not satisfactory for our
phenomenological purpose of being able to accommodate, 
for example, the Standard Model among their low-energy states.

Thus, the critical issue of whether there exist stable non-supersymmetric
strings {\it without}\/ boson/fermion degeneracies remains unknown.
However, our point is that if such a string is found, it will necessarily 
exhibit a misaligned supersymmetry as discussed above.  It will therefore
already incorporate solutions to the technical gauge hierarchy and 
cosmological constant problems simultaneously, in the sense described above.
If nothing else, this observation adds urgency to the search for a stable,
non-supersymmetric string.

\section{Discussion and open questions}
\setcounter{footnote}{0}

In this paper, we have proposed an alternative perspective concerning the
gauge hierarchy and cosmological constant problems.
Rather than address these problems through the language of a low-energy effective
field theory comprising a finite number of states, we have proposed an alternative
solution in which an infinite number of states at all energy scales
conspires to remove the quantum-mechanical sensitivity to high scales that would otherwise
appear to exist in the calculations of the Higgs mass and the cosmological constant.
As discussed above, the critical ingredient in this approach
is modular invariance and the misaligned supersymmetry that ensues in the
spectrum of physical string states.  As such, this proposal would be realized naturally
within the context of a stable, non-supersymmetric string. 
Indeed, one of the advantages of this approach is its generality:  
since our observations are built only on modular invariance and on the ensuing misaligned
supersymmetry, they apply to all (stable) non-supersymmetric closed strings 
regardless of their method  of construction or other phenomenological properties.

\subsection{Alternative approach to string phenomenology?}

\begin{figure}[ht]    
\vskip -2.2 truein
\centerline{
   \epsfxsize 6.7 truein \epsfbox {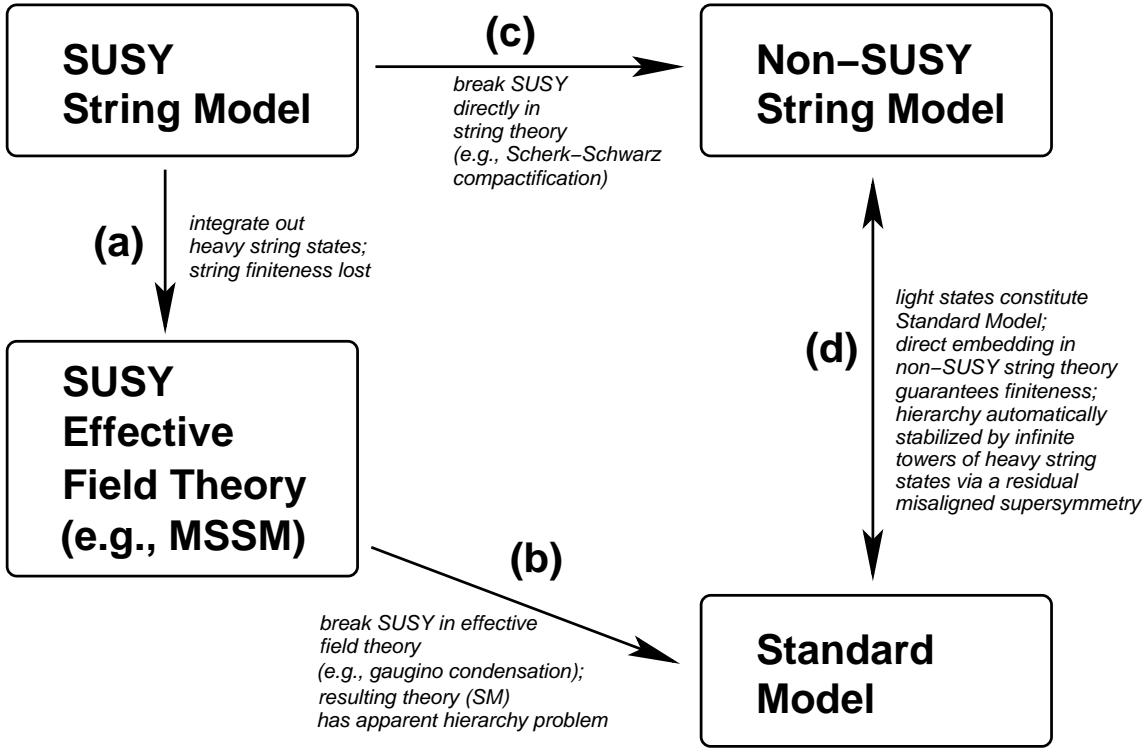}
    }
\vskip -2.2 truein
\caption{ Two approaches to string phenomenology.  In the traditional
         approach [paths~(a) and (b)], the Standard Model is realized
         only after supersymmetry is broken in an effective theory 
         (\eg, the MSSM) which is itself derived as the low-energy limit
         of a supersymmetric string.  In the alternative approach
         [paths~(c) and (d)], supersymmetry is broken first in a manner
         that preserves the full string symmetries that underlie string
         finiteness.  The light degrees of such a string theory would
         then constitute the Standard Model directly, and the gauge hierarchy 
         would automatically be stabilized through a misaligned 
          supersymmetry.} 
\label{newpath}
\end{figure}

Needless to say, assuming that a such a string exists,
the proposals in this paper favor a corresponding alternative approach to string 
phenomenology.  
As sketched in Fig.~\ref{newpath} [paths~(a) and (b)],
the traditional approach to string phenomenology~\cite{stringpheno} 
has always been 
to begin with a supersymmetric string model at the Planck scale,
and then essentially to 
integrate out the heavy string states, leaving behind
a supersymmetric effective field theory (\eg, the Minimal Supersymmetric Standard Model)
comprising only the light (or massless) string degrees of freedom.
As we have seen, this process of integrating out the heavy string states
eliminates the fundamental finiteness properties
that are intrinsic to the full string theory as a result of a conspiracy between the
states at all energy scales.
Then, as a second step, one typically breaks supersymmetry in this effective field theory through
some field-theoretic mechanism (\eg, gaugino condensation), ultimately 
resulting in the non-supersymmetric Standard Model.  It is this final theory
which has an apparent gauge hierarchy problem.  

By contrast, the ideas in this paper favor an alternative approach 
[indicated along paths~(c) and (d)]
in which supersymmetry is broken first~\cite{scherkschwarz,othernonsusy}.  
Unlike path~(a), such supersymmetry breaking
occurs directly within the context of the {\it full}\/ string theory through 
a mechanism (such as Scherk-Schwarz compactification~\cite{scherkschwarz}) 
which respects the full underlying string symmetries, including 
worldsheet modular invariance.\footnote{
       In this connection, we emphasize that we are discussing {\it worldsheet}\/
       modular invariance.  In the literature one often encounters discussions
       of modular-invariant supergravities and modular-invariant gaugino 
       condensation (see, \eg, Ref.~\cite{MKG}),
       but those discussions involve modular $SL(2,\IZ)$ symmetries of 
       fields in the {\it target space}\/.}
Assuming that such a stable non-supersymmetric string exists,
this procedure ensures that the original finiteness properties of the string
theory remain intact.  In such a scenario, the light states of this string would then 
constitute the Standard Model directly, and the direct embedding of the Standard
Model within such a  non-supersymmetric string would automatically stabilize the gauge
hierarchy in the manner we have been discussing.       
 
In principle, both of these approaches provide a connection between a string theory
at the Planck scale and the non-supersymmetric Standard Model at lower scales. 
However, the physics {\it beyond}\/ the Standard Model clearly depends on the route taken.
The traditional route [paths~(a) and (b)] requires low-energy (weak-scale) supersymmetry
in order to protect the gauge hierarchy,
and does not directly address the cosmological constant problem.
By contrast, the alternative route [paths~(c) and (d)] does not require supersymmetry at
any energy scale either at or below the string scale.  The gauge hierarchy and cosmological
constant problems are instead directly related to the conjectured stability of the
non-supersymmetric string via the misaligned supersymmetry that remains in the
full string spectrum.  

Of course, one interesting possibility may be that the diagram
in Fig.~\ref{newpath} actually {\it commutes}\/.
In other words, it is possible that after gaugino condensation [path~(b)], 
the states that emerge are still realizable as the low-energy states of a
non-supersymmetric string in which the scale of supersymmetry breaking is
somehow much smaller than the string scale.\footnote{Models 
       of this sort can be found in Ref.~\cite{Antoniadis}.}  
However, this would require that
the {\it field-theoretic}\/ mechanism of gaugino condensation somehow
preserve the full {\it string-theoretic}\/ symmetries 
(such as worldsheet modular invariance) that underlie string finiteness. 
Whether this is possible remains unclear.

\subsection{Relevance for the ``brane world''?}

Our approach may also be relevant for the so-called ``brane world'' scenario of
physics beyond the Standard Model~\cite{extradims,RS,recent}.
At first glance, our approach is a ``no-braner'', since the Standard Model
is not necessarily restricted to any particular subspace of the compactified string theory.
However, our approach may be useful even within brane-world scenarios 
with reduced GUT, Planck, and string scales.  Since our results apply
for closed strings,  they are applicable in
the ``bulk'' of extra large dimensions, where all degrees of freedom
are necessarily realized as excitations of closed strings.  
As such, it might be possible to follow this approach (rather than supersymmetry) 
in order to stabilize bulk dynamics and ensure finiteness in the bulk.  
However, it is possible that generalizations of our results would also apply  
in the open-string sectors of Type~I string models.  If so, then this approach
could be used to ensure finiteness on the brane as well as in the bulk, and permit
the fundamental string scale to exceed the TeV scale in such scenarios.
In this context, it would be interesting to study the relation between this
approach and the phenomenon of brane supersymmetry-breaking~\cite{branesusybreaking}.

Of course, the precise low-energy phenomenology 
of our scenario depends on the value of the string scale $\mu$.  {\it A priori}\/,
this scale is a free parameter in our approach, since the misaligned supersymmetry
and the cancellation of the mass supertraces 
holds regardless of the value of the fundamental scale $\mu$.
If $\mu$ is large (\eg, near the usual Planck scale), then at low energies
we would not expect to detect new physics other than the ``extra'' light states
(beyond the Standard Model) that generically emerge in quasi-realistic 
string models.
On the other hand, if $\mu$ is near the TeV range, then the full spectrum of
string states would become accessible to upcoming collider experiments, 
and the appearance of a misaligned supersymmetry 
should become experimentally verifiable.
Of course, $\mu$ may also take intermediate values~\cite{intermediate}.

\subsection{Open issues}

Needless to say, this approach raises a number of outstanding issues.
As such, the comments that follow are highly speculative.

First, as mentioned above, our specific results are proven
merely to one-loop order
for closed strings in a flat background spacetime.  
In other words, misaligned supersymmetry and the corresponding
supertrace relations such as those in (\ref{conc1}), (\ref{conc2}),
and (\ref{sutracerels}) hold for tree-level masses and one-loop cosmological
constants.
It is still necessary to demonstrate that misaligned supersymmetry
and the vanishing of the cosmological constant persist to all orders,
and perhaps even non-perturbatively. 
Of course, as discussed in Sect.~5, this is nothing but the usual moduli
problem.    However, the root of our approach is ultimately 
modular invariance, and this is a symmetry with well-understood multi-loop
generalizations.  It is therefore likely  that multi-loop generalizations 
of modular symmetry will continue to tie the hierarchy and
cosmological constant problems together, and relate them ultimately to the
over-riding (unsolved) problem of non-supersymmetric vacuum stability.

A second, closely related issue concerns the extension of these results to open
strings.  Although modular invariance can no longer be expected to hold, 
open strings can be realized as orientifolds of closed 
strings~\cite{orientifolds}.  As such,
remnants of misaligned supersymmetry may survive the orientifolding procedure. 
Indeed, one of the critical ingredients in our supertrace derivation is
the result (\ref{KSresult}), which rests directly on modular invariance.
However, this result has recently been generalized to open strings~\cite{KStypeI}.

Moreover, in many instances open strings emerge as the strong-coupling duals
of closed strings~\cite{PolWit}, and such strong/weak coupling duality relations 
have been conjectured to hold even
without supersymmetry~\cite{nonsusyduality}.
In such instances, a successful generalization of our results to open 
strings would therefore allow us to conclude that that our closed-string results would
hold even non-perturbatively.

A third open issue concerns the origin of the scale of electroweak symmetry
breaking.  Clearly, our approach does not shed light on this critical issue.
This scale is presumably set by some dynamics
connected with the stability of the underlying non-supersymmetric string.
However, regardless of how this scale is set, our point is that it is 
then guaranteed to be insensitive to other heavy scales such as the string scale.  
Thus, once this scale is set, 
the {\it technical}\/ hierarchy problem is solved, and any apparent
quantum-mechanical sensitivity to heavy scales is merely an artifact of having 
truncated our theory to include only the light degrees of freedom.

Similar comments may also apply to the cosmological constant problem.
Of course, our results assume the propagation of our strings on a flat background
spacetime;  in such cases the one-loop condition for vacuum stability 
becomes the vanishing of the one-loop cosmological constant.  In some sense,
this may be considered to be an automatic ``self-tuning'' since all other non-supersymmetric
strings are necessarily unstable (with non-zero dilaton tadpoles) and thus necessarily
flow in string moduli space until they reach a stable point.  
Of course, how this generalizes to higher loops and curved spaces is more 
difficult to study (especially since string theory undoubtedly demands more than
a semi-classical treatment of the background fields).  
However, it is likely that a true solution to the cosmological constant problem
will ultimately require physics that is intrinsically {\it stringy}\/,
and which therefore cannot be captured within a semi-classical treatment of background 
gravitational fields (alone or in the presence of branes).

Nevertheless, one interesting observation concerns phase transitions
in the early universe.
As discussed above, our assumption in this paper has been that 
our present-day low-energy world can be directly realized as
the low-energy states of a non-supersymmetric string.
If this assumption holds at each epoch in the evolution of the
universe, then all phase transitions should occur in such a way that they
do not violate fundamental string symmetries such as worldsheet modular 
invariance.  In other words, unless modular invariance is broken spontaneously,
modular invariance should continue to protect
the cancellations we have observed by causing the infinite towers of string
states to experience phase transitions and adjust their properties
whenever the low-energy states do the same.
Thus, the ``self-tuning'' described above would be maintained,
and modular invariance would be preserved {\it through}\/ all phase transitions,
including the QCD phase transition~\cite{cudell}.

A fourth outstanding issue concerns the relevance of effective field theory,
particularly as it relates to the phenomenological study of theories (such
as string theories) in which there are infinite numbers of heavy states.
Needless to say, our approach to the gauge hierarchy problem 
is one which goes against the spirit of effective field theory.  Unlike
other approaches to this problem (such as supersymmetry), our approach rests on 
conspiracies between physics at all energy scales simultaneously, and  
illustrates  the apparent fallacy of integrating
out {\it infinite}\/ numbers of heavy string states.
Or, phrased somewhat differently, the low-energy effective field theory
derived from the string is one in which certain parameters (\eg, supertraces) 
are magically cancelled by physics that cannot be captured within 
an approach based purely on low-energy physics.
It would therefore be interesting to clarify for which classes of phenomenological
problems in string theory an effective field theory approach is valid,  
and for which classes of problems a treatment within the framework of the full string theory
is required.  
Clearly problems involving relations between widely separated scales (such as
the gauge hierarchy and cosmological constant problems) are likely to be in 
the second category. 

In this connection, note that the cosmological constant problem is usually
considered to be an infrared problem.  By contrast, in string theory, there
is no distinction between ultraviolet and infrared physics, since  
modular invariance exchanges the two and provides a direct connection between
low-energy states and those which are infinitely heavy. 
It is therefore natural that our approach to the cosmological constant problem
is one which necessarily involves all energy scales simultaneously.
Similar remarks may also apply to solutions to the cosmological constant problem
which are based on the holographic principle~\cite{holography}.

Conversely, it may also be interesting to study the extent to which
modular invariance may be exploited as a new regularization mechanism entirely
within a field-theoretic context (albeit a field theory with an infinite number of
states).  
In some sense, one may think of modular invariance as providing an infinite-component Pauli-Villars
regulator in which infinite numbers of heavy fields cancel the divergences of light fields.
It would be interesting to explore the connection between this method of achieving
finiteness and 
finiteness achieved via soft supersymmetry breaking~\cite{JJ}. 
There may also be interesting connections
to other non-traditional field-theoretic regularization mechanisms such as ``non-local 
regularization''~\cite{regs} 
and ``Kaluza-Klein regularization''~\cite{KKreg}.
Indeed, misaligned supersymmetry may specifically be able to shed light on some of
the unsolved features involved in the latter.

The ideas we have proposed in this paper are clearly speculative.  
At the very least, they rest critically on the existence
of stable non-supersymmetric strings, a fact which has not yet been demonstrated.
However, our main point is that any such string theory must necessarily
exhibit the properties we have discussed, since modular invariance and misaligned
supersymmetry are intrinsic ingredients in the self-consistency of string theory.
As such, {\it this}\/ appears to be the path to finiteness 
chosen by string theory.  For this reason alone, we believe that this approach
merits further exploration.

\section*{Acknowledgments}

This work is supported in part by the National Science Foundation
under Grant~PHY-0007154,
and by a Research Innovation Award from Research Corporation.
I wish to thank
 S.A.~Abel,
 E.~Dudas,
 P.H.~Frampton,
 T.~Gherghetta,
 G.L.~Kane,
 S.F.~King,
 G.~Kribs,
 C.S.~Lam,
 J.~March-Russell,
 S.P.~Martin,
 M.~Moshe,
 J.~Moffat,
 R.C.~Myers,
 A.~Nelson,
 A.~Peet,
 M.~Porrati,
 H.~Tye,
 U.~van~Kolck,
 and J.~Wells
for discussions of these and related topics.
I also wish
to acknowledge the hospitality of the Aspen Center for Physics
where portions of this work were performed.

\bigskip
\vfill\eject

\bibliographystyle{unsrt}

\end{document}